\documentclass[twocolumn]{revtex4-1}
\usepackage{graphicx}
\usepackage{amsmath}
\usepackage{amssymb}
\usepackage{bm}
\usepackage{color}

\newcommand{\B} {\bm{B}}
\newcommand{\X} {\bm{X}}
\newcommand{\x} {\bm{x}}
\newcommand{\kk} {\bm{k}}

\newcommand{\rhoL} {\bm{\rho}}
\newcommand{\wx} {\varpi}
\newcommand{\wv} {\varpi}

\begin{document}

\title{Gyroaveraging operations using adaptive matrix operators}
\author{Julien Dominski}
\email{jdominsk@pppl.gov}
\author{Seung-Hoe Ku}
\author{Choong-Seock Chang}
\affiliation{Princeton Plasma Physics Laboratory, PO Box 451 Princeton, NJ 08543, USA}

\begin{abstract}
A new adaptive scheme to be used in Particle-In-Cell codes for carrying out gyroaveraging operations 
with matrices is presented. This new scheme uses an intermediate velocity grid whose 
resolution is adapted to the local 
thermal Larmor radius.  The charge density is computed by projecting marker weights in a field-line 
following manner while preserving the adiabatic magnetic moment $\mu$. 
These choices permit to improve the accuracy 
of the gyroaveraging operations performed with matrices even when strong spatial variation of temperature and magnetic field is 
present. Accuracy of the scheme in different geometries from simple 2D slab geometry 
to realistic 3D toroidal equilibrium has been studied. A successful implementation in the grokinetic code XGC is presented in the delta-f limit.
\end{abstract}

\maketitle

\section{Introduction}
Gyrokinetic codes are one of the best candidates for studying the turbulent transport of tokamak fusion plasma. 
These codes are based on the gyrokinetic theory which reduces the 6D Vlasov equation into a 5D gyrokinetic 
equation~\cite{Brizard07}. This reduction is made possible thanks to the fact that the fast gyromotion 
of a particle around a magnetic field line is associated to an adiabatic 
invariant, the magnetic moment $\mu$ . A particle is thus modeled by a ring of charge which 
radius is called the Larmor radius and center is called the gyrocenter. The electric field felt by 
this ring of charge is called the gyroaveraged electric field and is computed from the electrostatic 
potential averaged over the same ring. In the present work, we report a new scheme based on matrix operation 
to be used in PIC codes for computing this gyroaveraging operation in a fast and accurate manner. 

The gyroaveraging operation can be handled with two different strategies, using either matrix 
operations over grid quantities or a finite number of points over each individual marker particle gyro-orbit approximated as circle. 
When the quantity to be gyroaveraged is represented on a grid, as it is the case in Eulerian continuum codes, 
the operations can be computed with matrix operations. 
When the quantity to be gyroaveraged is sampled by marker particles, as it is the case   
in most Lagrangian Particle-in-Cell codes, the  operations in velocity space are handled for each 
marker particle separately. The gyroaveraging operation then consists in projecting the marker gyrocenter $\X$ on its discrete 
gyroring  $\x_a=\X+\rhoL(\alpha_a)$ with $n$ points labeled by $a$ and located at equidistant 
gyro angles $\alpha_a=a\,2\pi/n$. 
This operation can be computed with at least 4 points~\cite{Lee87} or more than 4 points according to 
the physics solved by the system~\cite{Mishchenko05}. 

The gyrokinetic code XGC has the particularity of being the only PIC 
code which uses gyroaveraging matrices. For this purpose, the particle weights are projected on an 
intermediate 4D grid on which the gyroaveraging is performed. This functionality 
was first implemented by S. Ku~\cite{Ku16,Ku18}. The current work reports on a new scheme 
in which the gyroaveraged matrices are adapted to the local thermal Larmor radius and the projection 
of particle weights is made such as to preserve their adiabatic moment $\mu$.  In general, gyrokinetic PIC codes like ORB5~\cite{Tran99}, 
GEM~\cite{Parker03}, or GT5D~\cite{Idomura08} use the $n$ points gyroaveraged 
technique~\cite{Lee87,Lin95}. On the other hand, Eulerian codes like GENE~\cite{Jenko00,Goerler11,Jarema17}, 
or the semi-Lagrangian code Gysela~\cite{Grandgirard07,Steiner15,Rozar16}, which are grid-based, 
can naturally use gyroaveraging matrices. 
The direct use of the Bessel function or of its Pad\'e approximation can also be made by these grid based codes, such as in Gyro~\cite{Candy03}. 

In the new scheme, the gyro-averaging matrix operations are performed on a 4D grid composed of the 3 dimensions 
of position space, $\bm{X}$, \textit{plus} a dimension in the velocity direction $\mu$. The choice of using a grid 
in $\mu$ instead of a grid in $\rho$ or $v_\perp^2$ is made because $\mu$ is an adiabatic invariant. 
The grid in the $\mu$ direction is regular in $\sqrt{\mu}$ and is referred to as a 
$\sqrt{\mu}$-grid.  This $\sqrt{\mu}$-grid is adapted at each node point to the 
spatial variation of temperature and magnetic field strength, \textit{i.e.}, the maximum value of the grid 
$\sqrt{\mu_{\rm max}}$ is a multiple of the square-root of the thermal magnetic moment $\mu_{\rm th}=T/2B$. 
Projecting gyrocenters in 4D space with a $\sqrt{\mu}$-grid thus preserves 
their magnetic moment. In the previous XGC version, the same fixed $\rho$ grid was used at all position space 
positions. As we will discuss in this paper, this previous approach 
shows two inconvenients. First, in case of strong spatial variation of temperature, more velocity grid points are 
necessary to converge the integral of gyroaveraged quantities. Second, in case of strong variation of the magnetic field strength 
in the parallel direction (cf NSTX or tight aspect ratio tokamaks), more planes are necessary in the 
toroidal directions for converging the results. 

As we will discuss in more details, the computational cost for gyroveraging or integrating over $\mu$ accurately 
increases with $k\rho$ which is the ratio between the Larmor radius and the physical wavelength of interest. 
In ion turbulent regime such as in the ion temperature gradient (ITG) regime one has $k\rho_i\simeq0.5$, 
so that $4$ gyropoints are enough~\cite{Lee87}, as well as, a few points in $\mu$~\cite{Lin95}. But when 
gradually including the light electron physics, this ratio starts to increase significantly and more gyropoints are necessary. 
For instance, when including the physics of passing electrons near mode rational surfaces, short ion scale physics of the order of 
$k_r\rho_i\simeq20$ has to be included for computing turbulent transport~\cite{dominski15} and many more gyropoints have to be used. 
For example, $\simeq20$ gyropoints were used in recent PIC simulations of TCV 
turbulence including drift-kinetic electrons and gyro-kinetic ions~\cite{dominski17}. 
Finally, the major challenge consists in simulating gyrokinetic ions and electrons in multi-scale physics where both ITG and 
electron temperature gradient (ETG) turbulence scales are accounted for, thus going from 
$k\rho_i\simeq0.5$ up to $k\rho_i\lesssim60$, see flux tube studies~\cite{Goerler11,Howard14,Maeyama15}. 
In the present work, the accuracy of gyroaveraging will be discussed for various regimes, in preparation of future high fidelity 
gyrokinetic simulations of turbulence plasma.

In section~\ref{sec:gyrokinetic}, the basic electrostatic gyrokinetic model relevant for this work is briefly introduced for defining terms. 
In section~\ref{sec:newscheme}, the new $\sqrt{\mu}$-grid based scheme is introduced together with the classical $n$ 
point averaging scheme. In section~\ref{sec:errorgyro}, the error made by the $n$ points gyroaveraging is recalled, see~\cite{Lee87}. 
In section~\ref{sec:2d}, the accuracy of the new gyroaveraging scheme is discussed 
in a simple 2D slab geometry. The new scheme is also compared to the classic n-points gyroaveraging scheme and to the fixed $\rho$-grid matrix scheme. 
In section~\ref{sec:3d}, the new scheme is described in 3D toroidal geometry, compared to other schemes, and 
successfully implemented in the delta-f version of the gyrokinetic code XGC. 
In section~\ref{sec:conclusion}, a conclusion is drawn. 

\section{Gyrokinetic model}
\label{sec:gyrokinetic}
In gyrokinetic codes, the plasma dynamics is modeled with a reduced 5D Vlasov-Maxwell system 
of equations. The species gyrocentre distribution function
$f$ is evolved according to the gyrokinetic equation
\[
\frac{df}{dt}=\frac{\partial f}{\partial t}+\dot{\X}\cdot\frac{\partial f}{\partial\X}+\dot{v}_\parallel\frac{\partial f}{\partial v_\parallel},
\]
with $\X$ the gyrocenter, $v_\parallel$ the velocity in the parallel direction, and $\dot{\mu}=0$.

As our goal is to present our new numerical scheme, 
we only consider the electrostatic limit of the gyrokinetic model.  In this electrostatic limit, the
equations of motion are given by
\begin{equation}
\begin{cases}
\dot\X=\frac{1}{B_{\parallel}^\star}\left(v_\parallel \bm{B}^\star-\mu\nabla B\times \bm{b}-
q\nabla\langle\phi\rangle_\alpha\times\bm{b}\right)\\
\dot v_\parallel =-\frac{\dot\X}{mv_\parallel}\cdot(\mu\nabla B-q\nabla\langle\phi\rangle_\alpha),
\end{cases}
\end{equation}
with $\B^*=\B+(m/q)v_\parallel\nabla\times\B$, $\bm{b}=\B/B$, and $\langle\phi\rangle_\alpha$ the 
gyroaveraged electrostatic potential defined by
\[
\langle\phi\rangle_\alpha(\X,\mu)=\oint d\alpha\, \phi(\X+\rhoL(\alpha,\X,\mu)),
\]
where the particle position $\x=\X+\rhoL$ is decomposed in its gyrocenter $\X$ and Larmor vector 
$\rhoL$. The Larmor vector is defined by
\[
\rhoL=q^{-1}\sqrt{2B\mu}(\cos{\alpha}\,\bm{e}_1-\sin{\alpha}\,\bm{e}_2),
\]
with $\alpha$ the gyroangle and $\{\bm{e}_1,\bm{e}_2\}$ an orthonormal basis in the plan perpendicular to the magnetic field $\bm{B}$. 
In the present work, the Larmor ring is assumed to lie on a poloidal plane of constant toroidal angle, \textit{i.e.}, 
$\rhoL\cdot\nabla\varphi=0$ as it is done in general by most PIC codes. Since the difference between 
$B_T$ and $B$ is on the order $(B_P/B)^2$. It is not difficult to project the tilted Larmor ring onto the 
${\bm{e}_1, \bm{e}_2}$ plane and model it as an ellipse. $B_P$ and $B_T$ are the poloidal and the toroidal 
components of $B$, respectively.

The gyrokinetic Poisson equation, given here in the long wavelength approximation, reads 
\[
\nabla_\perp\cdot\frac{q_i n_0}{m_iB^2}\nabla_\perp\phi-\frac{en_0}{T_{e0}}\left(\phi-\langle\phi\rangle_{FS}\right)=\bar{n}_i-\bar{n}_e^{\rm NA}
\]
where $\langle\phi\rangle_{FS}$ is the flux-surface averaged electrostatic field, ${\bar n}_i$ is the 
gyro-averaged ion gyro-center density, ${\bar n}_e^{NA} = \bar{n}_e -(en_0/T_{e0})(\phi -\langle\phi\rangle_{FS})$, and $\bar{n}$ 
is computed with
\begin{equation}
\bar{n}=\int_{-\infty}^{+\infty} dv_\parallel \int_0^{+\infty} d\mu \oint d\alpha \, \delta f(\x-\rhoL,v_\parallel,\mu).
\label{eq:n}
\end{equation}
In the previous equation, the relation $\delta f(\x-\rhoL,v_\parallel,\mu)=\int d\X \delta(\x-\X-\rhoL)\delta f(\X,v_\parallel,\mu)$ was used. 
For simplicity,   
in the present work, we consider the delta-f model in which the distribution function $f$ is split 
into the background part $f_0$ and the perturbation part $\delta f$, such that $f=f_0+\delta f$. $n_0$ and $T_0$ are the 
density and temperature of the background $f_0$.

\section{The new scheme}
\label{sec:newscheme}
In the new scheme, the gyroaveraging operations are 
performed with matrix operations instead of using the classical $n$-point technique. 
In the gyrokinetic model introduced in the previous section, two quantities involve the gyroaveraging.
They are the gyroaveraged electrostatic field $\langle\phi\rangle_\alpha$ and the right hand side of Poisson 
equation $\bar{n}$.  Since the particle density is calculated from the kinetic distribution function, the charge 
density $\bar{n}$ also involves a $\mu$-integral over a gyroaveraged quantity. 
This integral is discretized using a grid in the $\mu$-direction and  
one gyroaveraging matrix is used per grid point $\mu_k$ of the $\mu$-grid.   

Let us point out, that for consistency, the same discrete weighting operation 
has to be applied for deposing the charge of a particle on $\bar{n}$ and for computing the gyroaveraged 
self-consistent electric field $-\nabla\langle\phi\rangle_\alpha$ which is used in the equation of motion. 

\subsection{Classic $n$-points technique used in PIC codes}
In a PIC code, the phase-space is sampled with marker particles of weight $w_p$ 
at positions $(\X_p,v_{\parallel,p},\mu_p)$. The perturbed Klimontovich distribution function, represented by 
\[
\delta \tilde{f}(\X,v_\parallel,\mu)=\sum_p\,w_p\,\delta(\X-\X_p)\times\delta(v_\parallel-v_{\parallel,p})\,\delta(\mu-\mu_p),
\] 
is thus projected on the 3D configuration space grid $\x_g$ with a relation of the form
\begin{equation}
\bar{n}_g=\bar{n}(\x_g)=\sum_p\,w_p\,\frac{1}{n_\alpha}\sum_{a=1}^{n_\alpha}\mathcal{P}_g(\X_p+\rhoL_a).
\label{eq:barng}
\end{equation}
The $\mathcal{P}_g$ operator projects the weight of each particle gyropoint $\X_p+\rhoL_a$ on the grid nodes labeled $g$ 
and reads 
$\mathcal{P}_g(\X_p+\rhoL_a)=\wx_{pa}^g$ with $\wx_{pa}^g$ 
the projected weighting number. 
In practice, the cell on which each particle gyropoint lies is identified and the projection is made on its grid nodes. 
This density is then used to solve for the electrostatic potential $\phi$ using the gyrokinetic  
Poisson equation. For ensuring energy conservation, the same numerical scheme is used for estimating 
the gyroaveraged electric field which is used in the equations of motion, meaning that 
the same weights $\wx_{pa}^g$ are used for estimating
\[
\langle\phi\rangle_\alpha(\X_p)=\sum_g\frac{1}{n_\alpha}\sum_{a=1}^{n_\alpha} \wx_{pa}^g\, \phi(\X_g+\rhoL_a),
\]
and 
\[
\bar{n}(\x_g)=\sum_p w_p \frac{1}{n_\alpha}\sum_{a=1}^{n_\alpha}\wx_{pa}^g.
\]

Note that some PIC codes use finite-elements and the Galerkin projection 
technique, see reference~\cite{Fivaz98}. Some PIC code also use a double gyro-averaging technique~\cite{Lee87}.

\subsection{The new scheme based on matrix operations}
The new scheme is composed of three steps. First, the particles are projected on a 4D grid 
$(\X_g,\mu_k)$ with an operation of the form
\begin{equation}
n(\X_g,\mu_k)=\sum_p\,w_p\,\mathcal{P}_g(\X_p)\times\mathcal{P}_k^g(\mu_p), 
\label{eq:projection_Xmu}
\end{equation}
where $\mathcal{P}_k^g(\mu_p)$ projects the weight in the $\mu$ direction 
at spatial grid point $\x_g$ and reads $\mathcal{P}_k^g(\mu_p)=\wx_{k}^{gp}$ with $\wv_{k}^{gp}$ the weighting number. 
A $\mu$-grid regular in $\sqrt{\mu}$ is preferred, because it has a better sampling property at the thermal energy range. 
Second, for each perpendcular-velocity grid value $\mu_k$ the field is gyroaveraged with 
\[
\langle n\rangle_\alpha(\x_g,\mu_k)=\mathcal{G}_{gg^\prime}^w \,n(\X_{g^\prime},\mu_k)
\]
where $\mathcal{G}^k$ is a gyroaveraging matrix composed of elements
\begin{equation}
\mathcal{G}^k_{gg^\prime}= \frac{1}{n_\alpha}\sum_{a=1}^{n_\alpha}\mathcal{P}_g(\X_{g^\prime}+\rhoL^k_a).
\label{eq:gyromat}
\end{equation}
For example, if $|\rhoL^{k}|=0$ then $\mathcal{G}^{k}_{gg^\prime}=\delta_{gg^\prime}$ with $\rhoL^k_a=\rhoL(\X_{g^\prime},\mu_k,\alpha_a)$. 
Finally, the charge density is obtained by summing over the different perpendicular contributions, by doing
\[
\bar{n}_g=\sum_{k=1}^{n_\perp} \Delta\mu^{g}_{k}\,\langle n\rangle_\alpha(\x_g,\mu_k),
\]
where $\Delta\mu^{g}_{k}$ is the $\mu$ grid spacing at position $(\x_g,\mu_k)$. The regular thermal grid 
in $\sqrt{\mu}$ is composed of 
points $\sqrt{\mu}_k=k\sqrt{n_{\rm max}\mu_{\rm th}(\x_g)}/n_\mu$ with $ \mu_{\rm th}(\x_g)=T(\x_g)/2B(\x_g)$ the 
thermal magnetic moment and $n_{\rm max}$ a real number taken big enough with respect to the particle loading.

The present work has two original aspects. First, the gyroaveraging matrices are 
assembled on a $\sqrt{\mu}$-grid which is normalized to the variations of the thermal Larmor radius. 
Second, the marker weights are projected in space in a way which preserves their 
adiabatic moment $\mu$. For a given gyrocenter, it consists in projecting this gyrocenter 
in space on spatial grid points  
prior to estimate its Larmor radii at each spatial grid point $\x_g$  separately with 
\[
\rho(\x_g)=\sqrt{2m\mu/q^2B(\x_g)},
\]
because the Larmor radius of a gyrocenter varies in space with respect to the magnetic 
field strength, as $\mu$ is a gyrokinetic adiabatic invariant.

The same scheme is applied for estimating 
the gyroaveraged electric field than for estimating $\bar{n}$, meaning that 
the same weights $\wx_g$ and $\wv_k$ are used for estimating 
\[
\langle\phi\rangle_\alpha(\X_p,\mu_p)=\sum_g\sum_k \wx_g^p \wv_{k}^{gp} \langle\phi\rangle_\alpha(\X_g,\mu_k),
\]
and 
\[
n(\X_g,\mu_k)=\sum_p w_p \wx_g^p \wv_{k}^{gp}.
\]

Let us now illustrate how $\mathcal{P}_k^g(\mu_p)$ in Eq.~\eqref{eq:projection_Xmu} can be computed. 
For example, a particle projected in configuration space on the grid point $\x_g$ will then be projected to velocity space 
on the $\sqrt{\mu}$-grid index 
\[
k=\min\left\{{\rm floor}\left[n_k\sqrt{\mu_p/ n_{\rm max}\mu_{\rm th}(\x_g)}\right] ,n_k-1\right\}
\]
with the weight 
\[
\wv_{k}^{gp}=k+1-n_k\sqrt{\mu_p/ n_{\rm max}\mu_{\rm th}(\x_g)}
\] 
and on the $\sqrt{\mu}$-grid index $k+1$ with the weight $\wv_{k+1}^{gp}=1-\wv_{k}^{gp}$. 
Proceeding this way corresponds to project the 
$\sqrt{\mu_p}$ of each particle in space. 

In term of performance, the use of the matrix technique instead of the $n$-points technique for gyroaveraging could significantly speed up 
the simulation. Using the classical technique requires the computation of many gyropoints, as well as their deposition on the grid. In 
case of an unstructured mesh, as it is the case of XGC, the deposition on a grid cell can be very expensive.  If a very important number 
of particles per cell is used ($\simeq10$k per cell in XGC), it is thus interesting to 
only deposit the gyrocenter on the grid and to perform the gyroaveraging operation with matrix operations, because the matrices are 
assembled once at initialization.  Also, the matrix technique requires the assembly of many gyroaveraging matrices and more communications. This could become a limitation when a very dense grid is used or when the number of particle per cell per species is small ($\lesssim100$).

\section{Error estimate of the gyroaveraging operation}
\label{sec:errorgyro}
Considering a plane wave $\mathcal{A}(\x)=\hat{\mathcal{A}}_ke^{\imath\kk\cdot\x}$,
its gyroaverage can be expressed with a $J_0$ Bessel function following that
\begin{eqnarray}
\langle\mathcal{A}\rangle_\alpha(\X,\mu)
&=&\oint d\alpha\, \hat{\mathcal{A}}_ke^{\imath\kk\cdot(\X+\rhoL)} \notag\\
&=& \hat{\mathcal{A}}_ke^{\imath\kk\cdot\X}\oint d\alpha\,e^{\imath k\rho\sin{\alpha}} \notag\\
&=&\mathcal{A}(\X)J_0(k\rho) \notag,
\end{eqnarray}
with $\rho=q^{-1}\sqrt{2B\mu}$ the larmor radius and $k$ the wavevector.

In a PIC code, one does not use the $J_0$ Bessel function for gyroaveraging,
but a $n$-points averaging technique in the $\alpha$ direction defined by 
\[
\oint d\alpha\,\mathcal{A}(\X+\rhoL(\alpha)):=\frac{1}{n_\alpha}\sum_{j=1}^{n_\alpha} \mathcal{A}(\X+\rhoL(2\pi \,j/n_\alpha)),
\]
which corresponds to approximating the Bessel function with
\[
\tilde{J}_0(k\rho):=\sum_{n=-\infty}^{+\infty}e^{\imath k\rho \sin{2\pi\,j/n_\alpha}},
\]
where one employed $n_\alpha$ gyropoints. This $n$-points technique is explicitly 
used when projecting marker weights on density grid, see in Eq.~\eqref{eq:barng}, 
but is also used to assemble gyroaveraging matrices, see Eq.~\eqref{eq:gyromat}.

Following references~\cite{Lee87,Mishchenko05}, the error due to the $n_\alpha$-points
gyroaveraging technique applied on a plane wave field of wavevector $k$ can
be estimated as
\begin{equation}
error:=\tilde{J}_0-J_0=2\sum_{l=1}^{\infty}J_{ln_\alpha}(k\rho).
\label{eq:J0error}
\end{equation}

\begin{figure}
\begin{center}
\includegraphics[width=8cm]{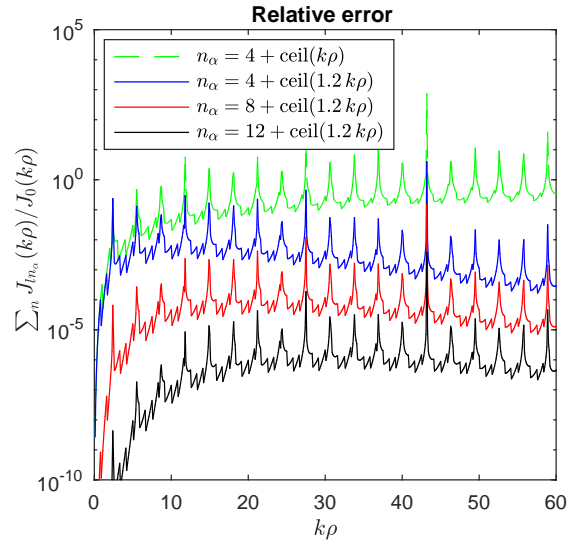}
\end{center}
\caption{Relative error of the discrete gyroaveraging operation, $(\tilde{J}_0-J_0)/J_0$, with respect to the product $k\rho$ 
for different number of gyropoints. For the blue curve, the number of gyropoints varies with $k\rho$ and is equal to $n_\alpha=4+{\rm ceil}(1.2k\rho)$ where ${\rm ceil}(x)$ is the ceiling function which provides the nearest integer bigger than the real number $x$.  
For example, ${\rm ceil}(\pi)=4$.}
\label{fig:choice_nalpha}
\end{figure}

Given this error estimate, a rule to ensure a certain accuracy of the gyroaveraging operation is 
defined. We plot some examples in figure~\ref{fig:choice_nalpha}. 
It is interesting to point out that the rule $n_\alpha=4+{\rm ceil}(1.2 k\rho)$ gives accurate results even for 
arbitrary big values of $k\rho$ which we tested with a simple code up to $k\rho\simeq100$. 
We observe that the number of points 
has to scale proportionally to $(1+\epsilon) k\rho$ with $\epsilon>0$ in order to keep the error bellow a certain level independently of 
the value of $k\rho$.  
The green curve, for which $\epsilon=0$, has an error increasing with $k\rho$. 
In comparison, the rule, using $\epsilon=0.2$, keep a relative error bellow a percent at all scanned values of $k\rho$ and 
could be considered for gyroaveraging the field felt by ion gyrocenters when including short scales physics 
of electrons down to ETG turbulence. Note that the peaks on these curves plotting the relative error are 
due to the zero of the Bessel functions $J_0$. 

\section{Numerical application of the new scheme in a 2D slab model}
\label{sec:2d}
In this section, we are interested in comparing the classical $n$-points gyroaveraging technique with 
grid-based techniques using gyroaveraging matrices. A simplified 2D slab geometry is considered, for which we know the 
analytical solution. 

Two techniques using the gyroaveraging matrices are presented: 
the \textit{fixed} $\rho$-grid technique which uses a fixed velocity grid at all position of the plasma 
and the \textit{adaptive} $\sqrt{\mu}$-grid technique which uses a thermal velocity grid adapted locally 
to $T$ and $B$ strength. One of the advantage of 
this new adaptive scheme is that its accuracy is independent of the variation of the thermal 
Larmor radius. Its correct usage requires some care when projecting the weights in order to preserve the 
gyrokinetic adiabatic invariant $\mu$. 

\subsection{Basic PIC model in 2D slab geometry}
A very simple 2D slab model is considered for studying the \textit{classical} and \textit{grid based} 
schemes used for gyroaveraging and integrating over the $\mu$-direction in a PIC code. 

The slab system consists of a box of lengths $l_x$ and $l_y$ in the radial and binormal 
directions, respectively. The magnetic field direction is perpendicular to the box. The perturbation wave is 
in the periodic $y$ direction and the temperature and magnetic field strength can vary in the $x$ direction. 
The grid is regular in $x$ and $y$ directions with respective intervals of length $\Delta x $ and $\Delta y$.

To represent a sinusoidal perturbation in this system, marker particles are loaded at random positions 
$(X_p,Y_p,\mu_p)$ with a weight $w_p=d\Omega \sin{(k Y_p)}$, where $p$ labels each particle quantities. 
Note that the magnetic moment of each particle might be overwritten to a given value according to the 
test we consider. Also, if we consider loading a mode for which $k=2\pi/\rho_{\rm th}$ then the 
SI value of this wave vector will vary with the temperature and the magnetic field strength, because 
$\rho_{\rm th}=\sqrt{mT}\,/qB$ so that $k=2\pi\sqrt{mT}/qB$ varies with $T$ and $B$. 

Let us now define the spatial projection, or the particle shape function, $\mathcal{P}_g$ which is used in this simple model to 
project the marker weights on the spatial grid points. The velocity space projection $\mathcal{P}_k^g$ was already 
defined in previous section~\ref{sec:newscheme}.

The projection of the gyrocenter, $\mathcal{P}_g$ in Eq.~\eqref{eq:projection_Xmu}, is computed by bi-linear interpolation. 
For each marker $p$ of weight $w_p$ and position $(X_p,Y_p)$, where $X_p=x_i+\epsilon_x\Delta x$ and $Y_p=y_j+\epsilon_y\Delta y$ 
with $\epsilon_x<1$ and $\epsilon_y<1$, the projection of its weight on spatial grid points is done according to the following equations
\begin{equation}
\begin{cases}
n_{i,j}=n_{i,j}      +\bar\epsilon_x \bar\epsilon_y w_p\\
n_{i+1,j}=n_{i+1,j} +\epsilon_x \bar\epsilon_y w_p\\
n_{i,j+1}=n_{i,j+1} +\bar\epsilon_x \epsilon_y w_p\\
n_{i+1,j+1}=n_{i+1,j+1}+\epsilon_x \epsilon_y w_p,\\
\end{cases}
\end{equation}
where $\bar{\epsilon}_{x,y}=(1-\epsilon_{x,y})$ and $\epsilon_{x,y}$ real positive numbers smaller or equal to unity. Given the definitions of section~\ref{sec:newscheme}, 
one identifies that on the node $\x_g=(x_i,y_j)$, one has $\wx_g^p\equiv \epsilon_x \epsilon_y$. 

The projection of the gyrocenter ring of charge, $\mathcal{P}_g(\rhoL_a)$ in Eq.~\eqref{eq:barng}, 
is also computed by bi-linear interpolation, by projecting each gyropoint used to represent the gyroring with the $n$-points technique. 
For each gyrocenter marker $p$ of weight $w_p$, each one of its 
gyropoints $x_{p,a}=X_p+\rho\cos(2\pi a/n_\alpha)=x_i+\epsilon_x\Delta x$ and $y_{p,a}=Y_p+\rho\sin(2\pi a/n_\alpha)=y_j+\epsilon_y\Delta y$ is projected on the spatial space grid points according to the following equations
\begin{equation}
\begin{cases}
\bar{n}_{i,j}=\bar{n}_{i,j}      +\bar\epsilon_x \bar\epsilon_y w_p/n_\alpha\\
\bar{n}_{i+1,j}=\bar{n}_{i+1,j} +\epsilon_x \bar\epsilon_y w_p/n_\alpha\\
\bar{n}_{i,j+1}=\bar{n}_{i,j+1} +\bar\epsilon_x \epsilon_y w_p/n_\alpha\\
\bar{n}_{i+1,j+1}=\bar{n}_{i+1,j+1}+\epsilon_x \epsilon_y w_p/n_\alpha.\\
\end{cases}
\end{equation}

\begin{figure}
\includegraphics[width=8.5cm]{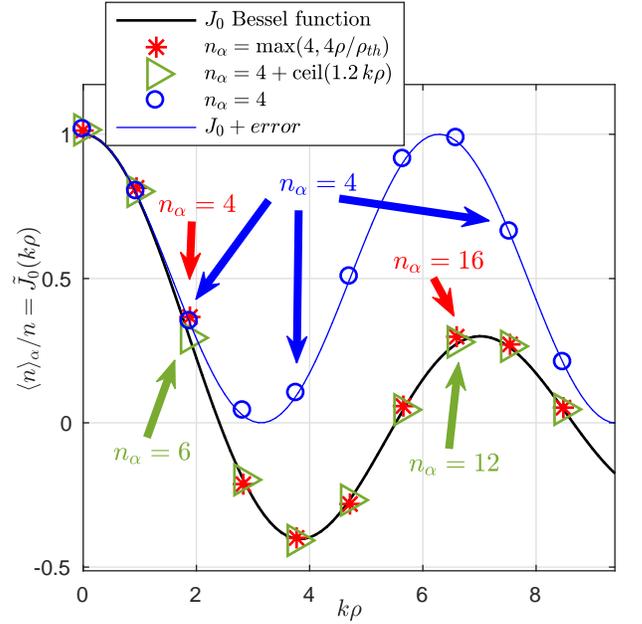}
\caption{Illustration of the gyroaveraging accuracy in a slab PIC model, by comparing the numerical Bessel factor 
$\tilde{J}_0=\langle n\rangle_\alpha/n$ (markers) and the solution $J_0$ (black curves). The blue curve is 
the analytical estimate of the numerical simulation results which consist of the sum of the solution $J_0$ and 
the $error$, see Eq~\eqref{eq:J0error}.}
\label{fig:gyroaverage_scalewithrho}
\end{figure}
\subsection{Accuracy of $n$-points gyroaveraging technique}

When gyroaveraging, 
the product $k\rho$ is the key parameter which permits to estimate how many gyropoints $n_\alpha$ on the gyroring 
are necessary. This is clearly shown by the equation
\[
\tilde{J}_0=J_0+2\sum_{l=1}^{\infty}J_{ln_\alpha}(k\rho)
\]
where the numerical error is composed of Bessel functions $J_{ln_\alpha}(k\rho)$. We thus choose to test this 
operation for values of the product $k\rho$ which are realistic for fusion plasma physics. The wavelength is set 
to $k\rho_{\rm th}=6\pi/10\simeq1.8$, which is relevant for trapped electron mode (TEM) instability. Different 
values of the particle Larmor radius are considered going from 
$\rho=0$ to $\rho=5\rho_{\rm th}$ which is a typical range of values with which marker particles are loaded in PIC codes 
for TEM mode simulation. In this case, he product $k\rho$ which requires the largest number of gyropoints will 
be the one used for gyrocenters loaded with $\rho=5\rho_{\rm th}$ and for which $k\rho=3\pi$. The size 
of their Larmor radius will be the biggest compared to the oscillation wavelength $2\pi/k$.

The test case is carried out for different values of $\mu\equiv\rho^2$. 
The two quantities $\mu$ and $\rho^2$ are exactly equivalent in this test case, because we consider 
homogeneous temperature $T$ and magnetic field strength $B$.  
The resolution is chosen big enough to accurately solve each mode and the  
box size is high enough to avoid any boundary condition issue. A simulation consists in loading 
markers in the box with a distribution chosen such that $n(\X)=\sin(kY)$ and to use the $n$-points 
gyroaveraging technique to compute $\langle n\rangle_\alpha(\x)$ and obtain a PIC estimate of the 
Bessel function $\tilde{J}_0(k\rho)=\langle n\rangle_\alpha/n$ with respect to the product $k\rho$.

Results of this first test are plotted in figure~\ref{fig:gyroaverage_scalewithrho}. 
For each scanned value of the Larmor radius $\rho$, the numerical
estimates $\tilde{J}_0=\langle n\rangle_\alpha/n$ of $J_0$ where $\langle n\rangle_\alpha$ is computed 
with the $n$-points technique is plotted with blue circles when using 
$4$ gyropoints ($n_\alpha=4$), with red asterisks when $n_\alpha=\max(4,{\rm ceil}(4\rho/\rho_{\rm th}))$, 
and with green triangles when $n_\alpha=4+{\rm ceil}(1.2\,k\rho)$. 
The analytical estimate $\tilde{J}_0=J_0+error$, see Eq.~\eqref{eq:J0error}, of the solution obtained when using only 4 gyropoints is plotted 
in blue to verify that the estimate of $\tilde{J}_0$ with particles is correct. 
It also shows that using a fixed number of $4$ points 
is not accurate and that the rule $n_\alpha=4+{\rm ceil}(1.2\,k\rho)$ is a better choice. Note that we 
verified the accuracy of this rule for values of $k\rho\simeq100$ which is of interest in view of performing 
ITG-ETG multiscale simulations. Moreover, for big values of $k\rho$, the rule $n_\alpha=4+{\rm ceil}(1.2\,k\rho)$ 
requires significantly less points than the method which consists in taking $4$ points for 
the thermal Larmor radius and scaling it linearly for bigger values of the Larmor radius.

In more constraining situations, one needs to choose the number of gyropoints when considering $k$ to be 
the Nyquist limit of the grid, $k_{\rm max}=\pi/L$. This could require particular
treatment in polar like mesh where the resolution of the grid diverges near the polar axis. 
See Ref.~\cite{dominski17} for an example where a Fourier filter is used near axis.

\subsection{Accuracy of $\mu$-integral of gyroaveraged quantity}

\begin{figure}
\begin{center}
\includegraphics[width=8.5cm]{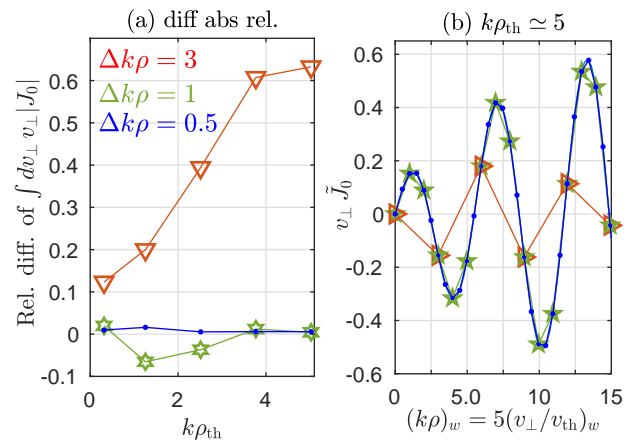}
\end{center}
\caption{Error made when evaluating the integral $\int d v_\perp \,v_\perp|J_0|$ with a discrete 
sum $\int dv_\perp=\sum_k \Delta _\perp$, as a function of the grid resolution $\Delta k\rho$. 
Subplot (a) shows the error estimate for different values of $\Delta k\rho$ with respect to $k$. 
Subplot(b) shows the approximation of the integrand $v_\perp J_0$ for different values of the grid size $\Delta k\rho$.}
\label{fig:study_vperp_integral_light}
\end{figure}

Performing the discrete $\mu$-integral of quantities gyroaveraged  with matrix operations 
corresponds to doing 
a quadrature in the $\mu$-direction over a grid. The accuracy of using $n_\mu$ gyroaveraging matrices 
is studied in this subsection. For this purpose, the error of the $\mu$-integration over the gyroaveraged quantities 
is analyzed when considering plane waves. The $\mu$-integral of such simplified 
problem would read
\begin{equation}
I(y)=\frac{B}{T}\int_0^{+\infty}d\mu\, J_0(k\rho) \sin(k y),
\label{eq:muintegral}
\end{equation}
where only $\rho$ depends on $\mu$. Given the definition of the Bessel function 
\[
J_0(k\rho)=\frac{1}{\pi}\int_0^\pi d\theta\,\exp{(\imath \,k\rho \, \cos(\theta))},
\]
one can appreciate the oscillatory nature of the $J_0$ function. On a plot of $J_0(k\rho)$, see for example 
Fig.~\ref{fig:study_vperp_integral_light}(b) where $v_\perp\tilde{J}_0$ is plotted, 
one could appreciate that the Bessel function oscillates similarly to a cosine of period $2\pi$. This indicates 
us that it is preferable to use a regular grid of points in $\rho$ or $\sqrt{\mu}$ rather than in $\mu$. 
Also, an accurate integral of a cosine function would need several points per oscillation 
period and at least two. This gives us a constraint on the number of quadrature points to  use:
$n_\mu> 2\max{|{\rm ceiling}(k\rho/2\pi)|}$. In practice, as we will show now, one needs 
$n_\mu> 6\max{|{\rm ceiling}(k\rho/2\pi)|}$.

The accuracy of using a set of gyroaveraging matrices to perform the $\mu$ quadrature is 
illustrated in our simple 2D slab model, see results in Fig.~\ref{fig:study_vperp_integral_light}. 
To measure the accuracy of this numerical integration, we measure the correctness of 
$\int d\mu|J_0|$. We study $|J_0|$ because $J_0$ is an oscillatory function whose integral 
is nearly zero and studying $|J_0|$ permits to better appreciate if the quadrature points 
reflect the structure of the function $J_0$ itself. The conclusion is that using $\Delta k\rho=3$ 
is clearly not precise enough, and using $\Delta k\rho=1$ or $0.5$ provides much more 
accurate results with an error of respectively a few percent or less, see subplot (a). 
The integrand $v_\perp\tilde{J}_0$ for the case $k\rho\simeq5$ is plotted in subplot (b) with the different grid resolutions. 
These results are obtained with our simple 2D PIC model. 

In PIC codes, there is in general no grid in $k\rho$, but a configuration space grid $\X_g$ and potentially a 
velocity grid $(\sqrt{\mu})_k$ or $\rho_k$. The max of $k\rho$ then depends on the max values of $k$ and $\rho$ separately. 
The max of $k$ should be at least the one of the dominant physical 
mode and at best the biggest value numerically solved by the grid at the Nyquist limit. 
For the velocity grid, one chooses the value which permits to include all loaded particles.

\begin{figure}
\begin{center}
\includegraphics[width=8.5cm]{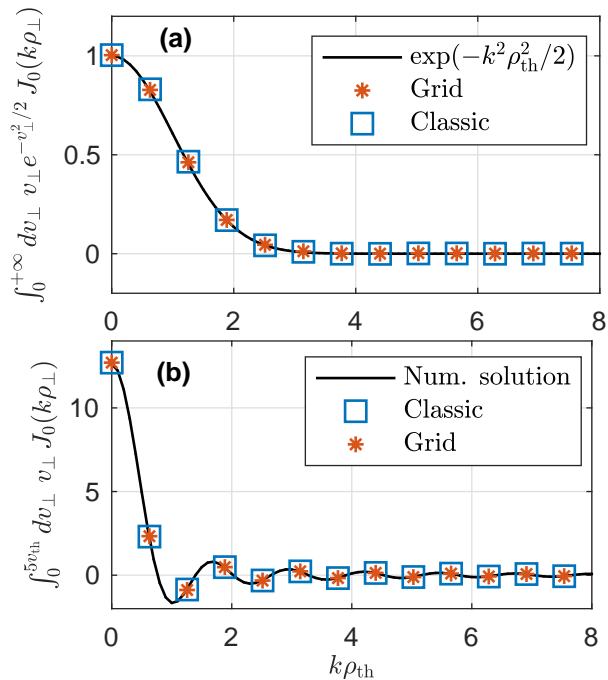}
\end{center}
\caption{Demonstration that using the classic $n$-points gyroaveraging technique and the new matrix based 
technique provide the same 
solution. The grid technique is using the new scheme with appropriate numbers 
of gyroaveraging points $n_\alpha$ and matrices $n_\mu$. These numbers where chosen 
according to the results presented in the text: $n_\alpha=4+{\rm ceil}(1.2k\rho)$ and $\Delta k\rho=0.5$. 
The test is carried out for various values of the perturbation wavevector $k$. Top subplot (a) 
is for integrating a Maxwellian background and bottom subplot (b) is when integrating a perturbation 
oscillation flat in $v_\perp$. }
\label{fig:Integrate_bessel_with_particles}
\end{figure}

We now perform an additional test for assessing the correctness of these operations performed on grids 
when using our two rules consisting in using a converged number of gyropoints 
($n_\alpha=4+{\rm ceil}(1.2k\rho)$) and of gyroaveraging 
matrices ($\Delta k\rho=0.5$). Results obtained with the 
classical integration and with the new scheme are compared in Fig.~\ref{fig:Integrate_bessel_with_particles}. 
As one can appreciate, the result obtained with the two methods agree very well. Note that 
this test is sensitive to the accuracy of the grid operations, such that the curves would deviate 
rapidly if not using enough points or matrices. 

\subsection{The new scheme based on the thermal grid}

The classical gyroaveraging technique is compared to the new grid-based technique, which uses a thermal $\sqrt{\mu}$-grid adapted locally 
to $T$ and $B$. For demonstrating the interest 
of the new scheme using adaptative matrices, we also consider 
the \textit{fixed} grid technique which uses the same $\rho$-grid at all positions of the plasma without following the variation of 
the thermal Larmor radius.  As will be shown, the new adaptive scheme accuracy is independent 
of the variation of the thermal Larmor radius. 

The main inconveniency of using a fixed $\rho$ grid is that 
the hottest and coldest part of the plasma have different thermal Larmor radius such that 
$\rho_{\rm cold}\ll\rho_{\rm hot}$. Therefore, when using the same $\rho$-grid of maximum 
value $\rho_{\rm max}$ and resolution $\Delta\rho=\rho_{\rm max}/n_\mu$, the maximum value will 
be dictated by the hottest part of the plasma and the resolution by the coldest part. 
For example, the biggest 
Larmor radius of the grid could be $\rho_{\rm max}\propto 5\rho_{\rm hot}$ and the resolution 
of the grid could be $\Delta \rho = \rho_{\rm cold}/2$. If one considers the extreme case 
for which $T_{\rm hot}/T_{\rm cold}=100$ and $k\rho_{\rm cold}=1$, as in the core-edge modeling of 
ITER with XGC1, then one needs $100=5*\sqrt{100}*2$ $\rho$-grid 
points with the fixed grid 
model, but one needs only $10=5*2$ $\sqrt{\mu}$-grid points with the new adaptive scheme 
in order to ensure the same accuracy. Note that the adaptive grid 
also accounts for the variation of magnetic field strength in both radial and poloidal directions, 
as we will discuss in section~\ref{sec:3d}.

The main idea for adapting the grid to the variation of temperature and magnetic field strength 
is to deposit the weight on a $\rho$-grid normalized to the local thermal Larmor radius. It 
consists in using a $\sqrt{\mu}$-grid which is adapted to the thermal Larmor radius at each 
configuration space grid point $\x_g$, where
\[
\tilde{\rho}_g=\rho(\x_g)/\rho_{\rm th}(\x_g)=\sqrt{\mu}/\sqrt{\mu_{\rm th}(\x_g)}=\sqrt{2\mu \,B(\x_g)/T(\x_g)}.
\]
The weight projected on the $\sqrt{\mu}$-grid point of index 
\[
k=\min(\tilde{\rho}_g\,n_\mu/\sqrt{n_{\rm max}} ,n_\mu-1)
\] 
at the node point $g$ would then be 
\[
\wv_k^{gp}=[(k+1)-\tilde{\rho}_g\,n_\mu/\sqrt{n_{\rm max}} ].
\] 
Proceeding this way corresponds to projecting the 
$\sqrt{\mu_p}$ of each particle in space; see section~\ref{sec:newscheme}.

\begin{figure}
\includegraphics[width=8.5cm]{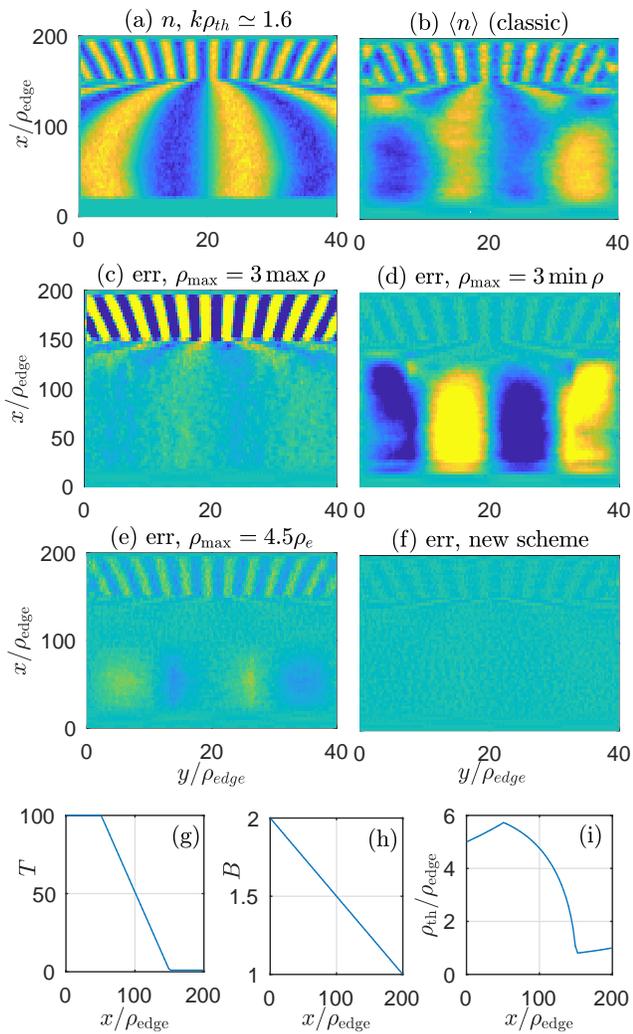}
\caption{Gyroaveraging with classic $n$-points or grid-based techniques in an extreme test case.  
Density to be gyroaveraged, $n(\X)=\sin(k(x)\,y)$ with  $k\rho_{\rm th}(x)=1.6$, is plotted in (a). 
Its gyroaveraged $\langle n \rangle_\alpha$ computed with the classic $n$-points technique  
is plotted in (b). The errors of the grid-based technique are plotted in subplots (c,d,e,f) where 
different choices for the grid in the $\mu$ (or $\rho$) direction have been made. In these four cases, 
$n_\mu=16$ matrices have been used, but 
the new scheme with thermal matrices is used in (f) and fixed grids are used in (c,d,e) with 
(c) $\rho_{\rm max}=3\max{\rho_{\rm th}}$, 
(d) $\rho_{\rm max}=3\min{\rho_{\rm th}}$, and (e) $\rho_{\rm max}=4.5\rho_{\rm edge}$. 
Radial profile of temperature (g), magnetic field strength (h), and thermal Larmor radius (i). $\rho_{\rm edge}$ is 
measured at $x/\rho_{\rm edge}=200$.}
\label{fig:2dslab_extrem_test}
\end{figure}
As an illustration, we consider another extreme case where there is a strong variation of the 
temperature profile, $T_{\rm max}/T_{\rm min}=100$, together with a variation of the magnetic field, see 
Fig.~\ref{fig:2dslab_extrem_test}(g) and (h). These variations lead to a more 
significant variation of the thermal Larmor radius (i), 
which permits to assess the interest of using the new adaptive scheme 
and its thermal $\sqrt{\mu}$-grid instead of using the fixed $\rho$-grid scheme. 
For this test, the gyrocenter perturbation which is gyroaveraged with the new techniques is 
$n(\X)=\sin(k(x)\,y)$ with  $k\rho_{\rm th}(x)=1.6$, see subplot (a). 
Its gyroaverage computed with the classical $n$-point technique is plotted in subplot (b). 
The errors of the grid-based technique are plotted in subplots (c,d,e,f) where 
different choices for the grid in the $\mu$ (or $\rho$) direction have been made. In these four cases, 
$n_\mu=16$ matrices have been used, but 
the new scheme with thermal matrices is used in (f) and fixed grids are used in (c,d,e) where 
for these thre later cases different values of $\rho_{\rm max}$ have been used: 
(c) $\rho_{\rm max}=3\max{\rho_{\rm th}}$, 
(d) $\rho_{\rm max}=3\min{\rho_{\rm th}}$, and (e) $\rho_{\rm max}=4.5\rho_{\rm edge}$. 
Despite using the same number of gyropoints $n_\alpha$ and matrices $n_\mu$, the errors made 
with the fixed grid technique, plotted in (c,d,e), are clearly bigger than  (f) the error made when 
using the new adaptive technique. 

\begin{figure}
\includegraphics[width=8.5cm]{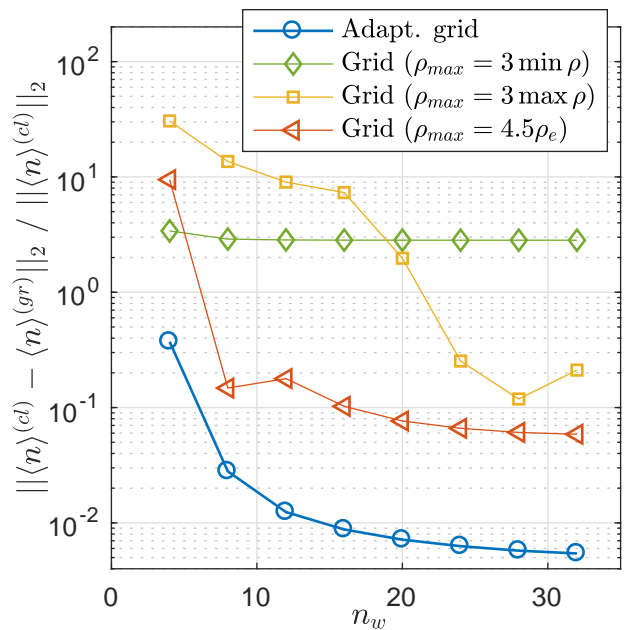}
\caption{Scan with respect to the number of gyroaveraging matrices of 
the difference between results obtained with the classic $n$-points gyroaveraged technique 
and with grid techniques. Same case than in figure~\ref{fig:2dslab_extrem_test}. 
Let us define the norm $|| \bar{n}||_2=\int_0^{L_x} dx \int_0^{L_y}dy \,(\bar{n}(x,y))^2$.}
\label{fig:2dslab_extrem_test_results}
\end{figure}
To further assess the accuracy of the technique using adaptive thermal matrices, a
scan in the number of matrices $n_\mu$  is carried out for this extreme test case in Fig.~\ref{fig:2dslab_extrem_test_results}. 
The error is estimated by integrating the difference with the expected $n$-points averaging solution 
\[
error=||\langle n\rangle^{(cl)}-\langle n\rangle^{(gr)}||_2/||\langle n\rangle^{(cl)}||_2
\]
where $\langle n\rangle^{(cl)}$ the field gyroaveraged with classic n-points technique 
and $\langle n\rangle^{(gr)}$ the field gyroaveraged with grid technique. The norm is defined by
$|| \bar{n}||_2=\int_0^{L_x} dx \int_0^{L_y}dy \,(\bar{n}(x,y))^2$. 
The accuracy of the new adaptive scheme 
is always better than the one of the fixed grid scheme. It is also shown than 
using a too small $\rho_{\rm max}$ will always lead to an incorrect solution independently of the number 
of matrices (green curve).

\section{New adaptive scheme in 3D toroidal geometry and its application to the gyrokinetic code XGC}
\label{sec:3d}
Compared to the simple 2D study of previous section, in 3D one needs to perform an additional 
projection of the particle weight in space. This additional projection is performed 
along the magnetic field line in a field-line-following manner. In the new adaptive scheme, the variation 
of the magnetic field strength along the magnetic field line 
is thus taken into account by projecting the $\sqrt{\mu}$ instead of estimating $\rho$ prior to project in space. 
A different $\sqrt{\mu}$-grid is used at each node point of the grid.

\subsection{New adaptive scheme in 3D toroidal geometry}
\begin{figure}
\begin{center}
\includegraphics[width=8.5cm]{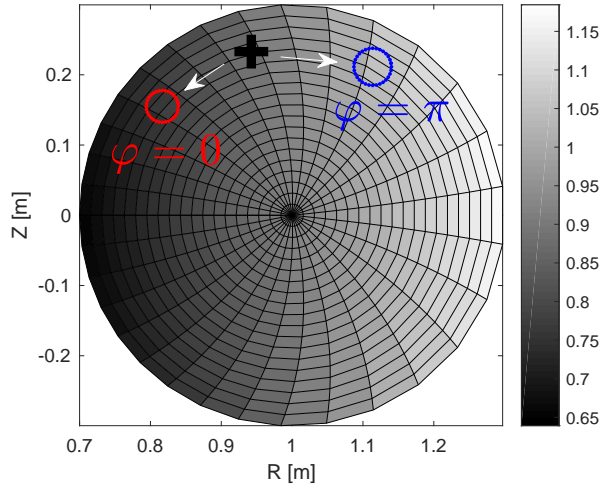}
\end{center}
\caption{Value of the thermal Larmor radius in an analytical \textit{ad-hoc} torus.  
$\rho_{\rm th}$ is computed for a constant temperature, such that in this case only the magnetic strength variation influences its value. 
The plotted grid is in the radius $r$ and the straight-field line angle $\chi$ coordinates. The black cross represents a guiding-center 
position at the poloidal plane $\Delta\phi/2$ and the red and blue rings are the projection 
of its Larmor rings on the poloidal planes $0$ and $\Delta\varphi$, respectively. }
\label{fig:larmor_projection}
\end{figure}
When projecting a particle (or gyrocenter) in the direction  of the magnetic field line on a surface $\psi$, its 
Larmor radius
\[
\rho=\sqrt{2m\mu/q^2B(\psi,\theta)}
\]
varies because of the variation of the magnetic field strength in the poloidal angle $\theta$. 
As an illustration, the Larmor radii of a given gyrocenter projected on 
two different planes in a field-following manner 
are plotted in figure~\ref{fig:larmor_projection}. The gyrocenter is the black cross on the plane $\varphi=\pi/2$, 
the red ring is the Larmor radius of its gyrocenter projected on the plane $\varphi=0$, and 
the blue ring is the Larmor radius of its gyrocenter projected on the plane $\varphi=\pi$. The strength of the magnetic field 
is plotted with a grey scale color code. 
The new scheme accounts for the variation of the Larmor radius, because it uses grid regular in $\sqrt{\mu}$ 
and $\mu$ is constant in space. This $\sqrt{\mu}$-grid is adapted at each node point $\x_g=(\psi_g,\theta_g)$, 
by setting its maximum value to 
\begin{equation}
\sqrt{\mu_{\rm max}(\x_g)}=\sqrt{n_{\rm max}\mu_{\rm th}(\x_g)},
\end{equation}
with $\mu_{\rm th}=T(\psi_g)/2B(\psi_g,\theta_g)$ the thermal magnetic moment and $n_{\rm max}$ an integer. 
The parameter $n_{\rm max}$ is chosen according to the loading of the particle, typical of a PIC code, such that 
for all marker particles, $p$, present in the vicinity of the node point $\x_g$, 
one has $\mu_p\le n_{\rm max}\mu_{\rm th}(\x_g)$.

A simple test case in 3D toroidal geometry is considered in order to illustrate that using an 
adaptive $\sqrt{\mu}$-grid is more accurate than using a fixed $\rho$-grid. 
We re-use the same simple PIC model of previous section extended in 3D.
The magnetic field equilibrium is chosen to be 
the circular \textit{ad-hoc} geometry which has a practical analytical definition as described in 
reference~\cite{Lapillonne}. The simulation volume is the 3D grid volume 
$L_r\times L_\chi\times L_\varphi$ with $L_r=a$ and $a=0.3m$ the minor radius, 
$L_\chi=2\pi$ in the periodic poloidal direction of the straight field line angle 
\[
\chi(r,\theta)=\frac{1}{q}\int_0^\theta B\cdot\nabla\varphi/B\cdot\nabla \theta , 
\]
 and $L_\varphi=2\pi/N$ is a fraction of the toroidal direction. The number of grid points 
 is $n_r\times n_\chi\times n_\varphi$ with $n_\varphi=2$, because we study the projection 
 of the marker particles on the planes located at $\varphi=0$ and $\varphi=L_Z$. 
 The volume in velocity space $L_\mu\times L_\alpha$ is represented with a grid of size 
 $n_\mu\times n_\alpha$.  The new scheme grid in the $\mu$-direction is regular in $\sqrt{\mu}$ 
 with $\mu_{\rm max}$ a varying quantity adapted to the temperature and magnetic field strength. 
The corresponding grid in the fixed grid scheme is regular in $\rho$ with $\rho_{\rm max}$ 
 a fixed value in all the simulation volume. For this test case, the position of the gyropoints are computed analytically 
 in the $(R,Z)=(r\cos\theta,r\sin\theta)$ plane 
 from the mapping between the geometrical angle $\theta$ and the straight-field line angle $\chi$
 \[
 \chi=2\arctan{\left[\sqrt{\frac{1-r/R_0}{1+r/R_0}}\tan{\left(\frac{\theta}{2}\right)}\right]},
 \]
 whose mapping is valid in the particular \textit{ad-hoc} geometry. $r$ is the minor 
 radius, $R_0$ is the major radius at magnetic axis, $R$ is the horizontal distance, and $Z$ the vertical distance.  
All markers are loaded between the poloidal planes $\varphi=0$ and $\varphi=L_\varphi=\Delta\varphi$. 

To compute the density $\bar{n}$ on each of these two planes, the marker particles are projected 
along the magnetic field lines. Given a marker $p$ with attributes $(r_p,\chi_p,\varphi_p,\mu_p)$, 
its projected position on the poloidal $\varphi=0$ plane is $(r_p,\chi_p-\varphi_p/q_s(r_p),0,\mu_p)$ 
and is $(r_p,\chi_p-(\Delta\varphi-\varphi_p)/q_s(r_p),0,\mu_p)$ on the $\varphi=\Delta\varphi$ plane, 
with $q_s(r_p)$ the safety factor 
\[
q_s=\frac{1}{2\pi}\oint d\theta \frac{\mathbf{B}\cdot\nabla\varphi}{\mathbf{B}\cdot\nabla\chi}.
\]
In this {\it ad-hoc} geometry, the surfaces have a circular cross-section such that the radius $r$ 
is a flux surface quantity, $r=r(\psi)$.

 In this section test case, the temperature is constant in the radial direction, 
 because we only want to study how the new scheme adapts to the magnetic field  
 variations in both radial and especially poloidal directions. 
 
\begin{figure*}
 \begin{center}
\includegraphics[width=12cm]{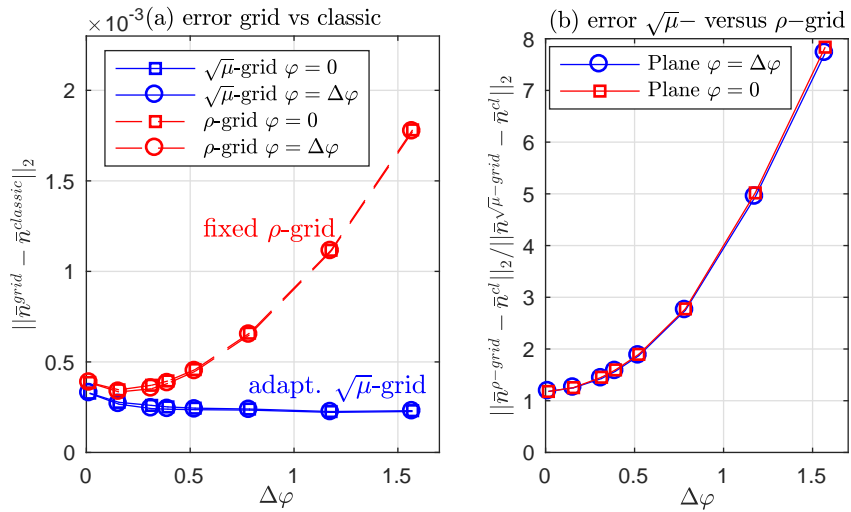}
\end{center}
\caption{Accuracy of gyroaveraging and $\mu$-integrals in 3D toroidal \textit{ad-hoc} geometry, estimated 
when computing  the guiding-center density $\bar{n}$, Eq.~\eqref{eq:n}. 
Plotted in (a) is the error made on $\bar{n}$ by grid techniques compared to classical $n$-points technique 
with respect to the distance, $\Delta \varphi$, between the poloidal planes on which the density is projected. 
Plotted in (b) is the ratio between the error made by the scheme using a $\rho$ projection on a fixed global velocity grid 
with the error made by the scheme projecting $\sqrt{\mu}$ on an adaptive local grid. In the latter, $\mu_{\rm max}$ of the 
grid is adapted to $T(\psi)$ and $B(\psi,\theta)$.
}
\label{fig:3d_matlab_scan}
\end{figure*}

 The perturbation is field-aligned in the form 
 \[
 \delta f = g(\mu)\, \sin{(q_s(r)\chi-\varphi)},
 \]
 where we arbitrarily took $g(\mu)=\delta(\mu-4\mu_{\rm th})$ for the test case of 
 figure~\ref{fig:3d_matlab_scan}. This choice is made for $g(\mu)$ in order to prevent that the 
 error made by various values of $\mu$ will compensate each others. Loading only one $\mu$ 
 permits to better compare the correctness of its projection on the velocity grid by the 
 different techniques. 

 Results are plotted in figure~\ref{fig:3d_matlab_scan}, where it is shown that the new scheme 
 using an adaptative $\sqrt{\mu}$-grid is more accurate than the scheme using a fixed $\rho$-grid. 
 In subplot (a), one can see (blue curve) that the new scheme is not 
 particularly sensitive to the distance $\Delta\varphi$ between consecutive poloidal planes. 
 On the contrary, the error of the scheme which  
 projects $\rho$ on a fixed $\rho$-grid is increasing significantly with $\Delta\varphi$. This $\rho$-grid 
 scheme is systematically less accurate than the new $\sqrt{\mu}$-grid scheme as shown in subplot (b) where the ratio 
 between the error of the fixed $\rho$-grid scheme with the error of the adaptative $\sqrt{\mu}$ scheme is plotted. 
 It shows that the new scheme is particularly effective for scenarios in which the magnetic field strength varies 
 strongly in the poloidal direction, which is the case of certain magnetic 
 configurations, such as the spherical tokamak MAST and NSTX.

\subsection{Application to the gyrokinetic code XGC}
\begin{figure}
\includegraphics[width=8.5cm]{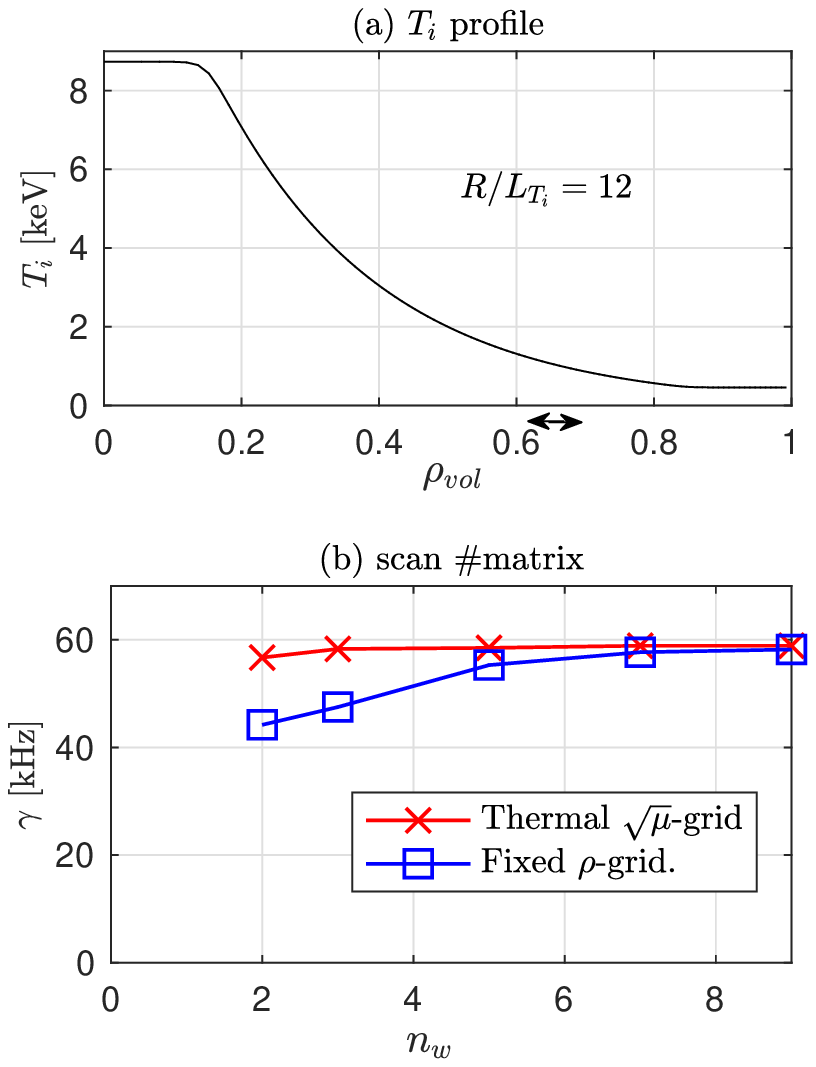}
\caption{Linear XGC delta-f simulation, convergence study of the growth-rate with respect to the 
number of gyroaveraged matrices $n_\mu$. Subplot (a) is the radial temperature profile. 
The temperature at axis is about $10$ times the temperature 
where the linear $n=64$ mode is growing (region identified with horizontal black arrows). $\rho_{vol}=\sqrt{V(\psi)/V(\psi_e)}$. 
Subplot (b) shows the values of the linear growth-rate of an $n=64$ mode 
$(k_\theta\rho_{\rm th}\simeq0.8$) for different values of the number of 
gyroaveraging matrices $n_\mu$.  Results from the simulation carried out with the fixed 
$\rho$-grid  and $\rho_{\rm max}=3.5\times\rho_{\rm th, axis}\simeq 2cm$ are in blue. Results 
obtained with the new adaptive grid and $\rho_{\rm max}(\psi,\theta)=3.5\times\rho_{\rm th}(\psi,\theta)$ 
are in red.}
\label{fig:lineartest_XGC_test_NEW_matrices_n64}
\end{figure}
To demonstrate the interest of using the new adaptive scheme in the gyrokinetic code XGC, we design   
a test case in which the fixed grid scheme will be at a disadvantage 
compared to the new adaptive scheme. For this purpose, we consider a 
case designed so that an ITG instability grows near the edge at a 
radial position $r/a\simeq0.8$  where the temperature is $10$ times smaller than the temperature 
at axis where $T_{axis}=8800$eV, see Fig.~\ref{fig:lineartest_XGC_test_NEW_matrices_n64}(a), 
and we build the fixed $\rho$-grid with a value of $\rho_{\rm max}$ equal to $3.5$ 
times the thermal Larmor radius at axis. This way, the resolution of the $\rho$-grid is not optimized for 
resolving the physics at $r/a\simeq0.8$. Indeed, $\rho_{\rm max}/\rho_{\rm th}(r/a\simeq0.8)\simeq10$ 
such that, having $k_\theta\rho_{\rm th}\simeq0.8$, one has $k_\theta\rho_{\rm max}=8$ and needs 
$n_\mu=7$ matrices to have $\Delta \rho k\simeq1$. This $\Delta k\rho$ resolution is necessary 
to perform accurate discrete integrals according to our 2D slab study of section~\ref{sec:2d}. 
In comparison, when using the adaptive $\sqrt{\mu}$-grid scheme, one has $\rho_{\rm max}=3.5\rho_{\rm th}$ and 
$k_\theta\rho_{\rm max}\simeq2.8$, such that $3$ matrices are enough for having  $\Delta k\rho\simeq1$.  
This analysis is confirmed by simulation results plotted in figure~\ref{fig:lineartest_XGC_test_NEW_matrices_n64}(b) where 
the growth-rate computed in the new adaptive grid scheme is converged at $n_\mu=3$ and  
the growth-rate computed in the fixed grid scheme is converged at $n_\mu=7$. 
The $\rho_{\rm max}$ is chosen $3.5$ times the thermal Larmor radius, because we load particles up to this value. 
The magnetic equilibrium is an ideal MHD equilibrium and corresponds to the geometry $5$ 
of reference~\cite{Burckel10}. Ions are gyrokinetic and electrons are adiabatic, for this test. 

\begin{figure}
\begin{center}
\includegraphics[width=8.5cm]{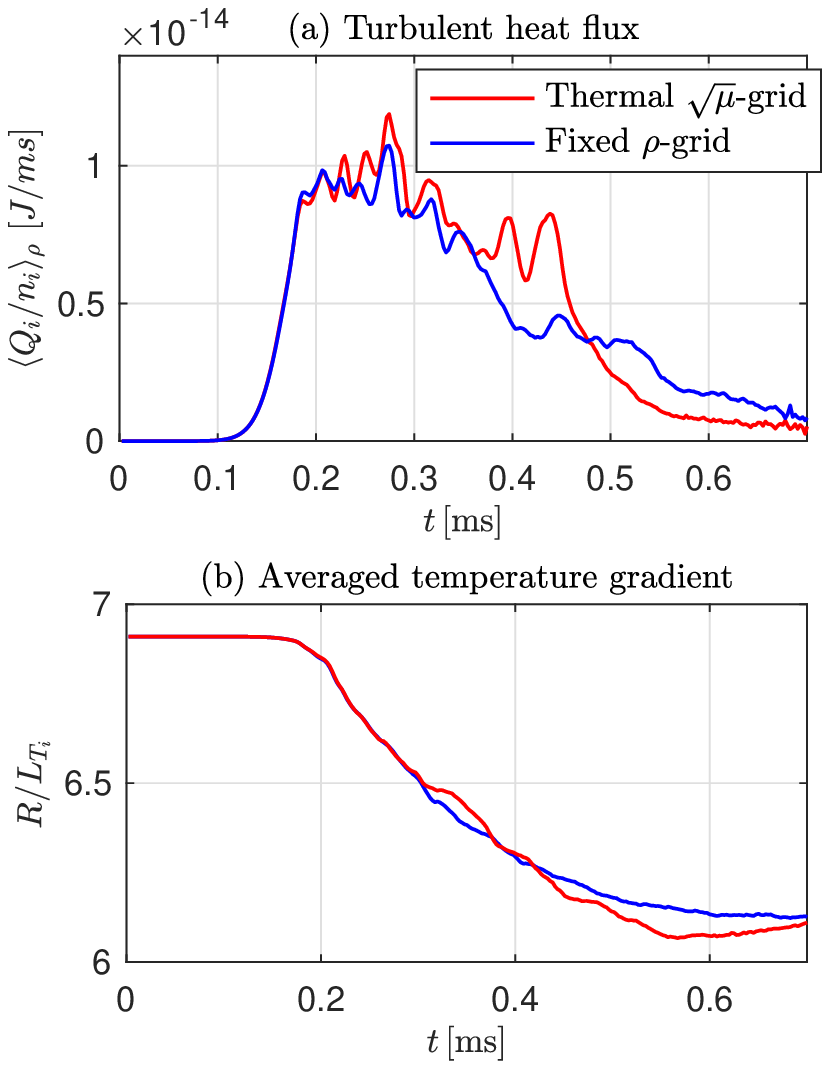}
\end{center}
\caption{Verification exercise. Nonlinear XGC simulation of a relaxation problem where the turbulence develop and the gradient 
of temperature is let free to relax. No heat source is employed. Results of simulations carried out 
with the old fixed $\rho$-grid scheme (blue) and with the new $\sqrt{\mu}$-grid scheme (red). 
Both simulations employ converged parameters for the gyroaveraging grid based techniques. 
This verification exercise show that the new matrix scheme does not inject an error.
}
\label{fig:nonlinear_case5_comparison_122M}
\end{figure}
A second nonlinear test is carried out  in order to ensure that no 
unforeseen error is caused by this new scheme. 
Simulations using either the fixed $\rho$-grid scheme or the adaptive $\sqrt{\mu}$-grid scheme 
are carried out with converged parameters and compared. 
This test is a nonlinear relaxation problem of an ITG regime 
in the same conditions as the previous case except that $R/L_{Ti}=6.9$ and that it 
is nonlinear. Typically, the plasma produces a strong flux of heat until 
its profile of ion temperature relaxes enough so that no turbulence is growing. Results are 
plotted in figure~\ref{fig:nonlinear_case5_comparison_122M}. In subplot (a), both simulations heat fluxes 
are in excellent quantitative agreement prior entering the turbulent saturated regime near $t\simeq0.19$ms. 
The deviation  occurring at latter times is essentially due to chaotic behavior of the turbulent regime. 
In subplot (b) the temperature relaxes in both simulations toward the same value $R/L_{T_i}\simeq6.1$ 
and its time evolution is very similar in both simulations.

\section{Conclusion}
\label{sec:conclusion}
A new scheme, based on the use of locally-adaptive gyroaveraging matrices for computing the gyroaverage of a field 
in gyrokinetic PIC code has been introduced. 
This new scheme permits to take into account the spatial variation of temperature and 
magnetic field strength in an efficient way. It also preserves the adiabatic moment $\mu$ 
when projecting a particle gyroring on the 4D grid. 
This new scheme has been studied in simplified 2D and 3D PIC models and implemented in the XGC code. 

The new scheme is based on a thermal 
grid in $\rho$, such that its accuracy is independent of 
these variation of temperature and magnetic field strength. As we discussed, using a thermal $\rho$ grid corresponds 
to using a $\sqrt{\mu}$-grid. In this work rules have been provided for choosing 
the resolution of the $\sqrt{\mu}$-grid, as well as for choosing the number of gyroaveraging  points. 
These rules have been illustrated with a basic PIC model when averaging the 
plane waves going from long wavelength ITG to short wavelength ETG. 

The accuracy of the $n$-points gyroaveraging technique and of the discrete integration over the velocity 
direction $\mu$ have been 
studied in slab geometry using a simple 2D PIC model. The product $k\rho$, which corresponds to the ratio of 
the Larmor radius with the physical wavelength, is the key parameter to consider 
when choosing both the number of gyro-points and the number of gyroaveraging matrices. 
The maximum value of the wavevector $k$ depends on the studied physics and on the Nyquist limit 
of the configuration space grid. The maximum value of $\rho$ depends on the loading of particles and on 
the local temperature and magnetic field strength. 

The importance of preserving the adiabatic moment $\mu$ when projecting the marker particle 
weights in configuration space has been shown by using a simple 3D PIC model. When projecting 
a marker in space, one must compute the Larmor radius at the projected position, because $\mu$ 
is an invariant but not $\rho$. 
Fusion plasma turbulence being anisotropic $k_\parallel\ll k_\perp$, the grid resolution is in 
general coarser in the parallel direction and finer in the perpendicular direction. 
Weights are thus projected over a longer distance in the parallel 
direction than in the perpendicular direction. This  projection over long parallel distances can lead to 
a significant variation of the Larmor radius with respect to the 
variation of the magnetic field strength in the poloidal direction. Projecting on an adaptive 
$\sqrt{\mu}$-grid permits to better account for this variation of the magnetic field strength. 
This feature is of particular interest in case of 3D magnetic equilibria 
in which the magnetic field strength varies significantly on the same magnetic surface, 
as it is the case of spherical tokamak or stellarator.
 
Finally, the new scheme has been successfully implemented in the gyrokinetic code XGC. 
To demonstrate its effectiveness, this new scheme using a thermal $\sqrt{\mu}$ grid has been 
compared to the scheme using a fixed $\rho$-grid. This interest has been shown in a case 
where the ion temperature varies significantly from the core to the edge, by a factor $10$. 
As expected from our preliminary studies, the new scheme requires much less point in the $\mu$ direction, which 
corresponds to less gyroaveraging matrices, 
for converging the growth-rate of the tested mode. 
The new scheme has also been verified in a simple nonlinear 
simulations including gyrokinetic ions and adiabatic electrons.

\section{Acknowledgement}
This research was supported by the SciDac project ``High-fidelity Boundary Plasma Simulation'', and 
the Exascale Computing Project (17-SC-20-SC), a collaborative effort of the U.S. Department of Energy 
Office of Science and the National Nuclear Security Administration.

This research used resources of the Oak Ridge Leadership Computing Facility at the Oak Ridge National Laboratory, which is supported by the Office of Science of the U.S. Department of Energy under Contract No. DE-AC05-00OR22725.

This research used resources of the National Energy Research Scientific Computing Center, a DOE Office of Science User Facility supported by the Office of Science of the U.S. Department of Energy under Contract No. DE-AC02-05CH11231.

This manuscript is based upon work supported by the U.S. Department of Energy, Office of Science, Office of Fusion Energy Sciences, and has been authored by Princeton University under Contract Number DE-AC02-09CH11466 with the U.S. Department of Energy. The publisher, by accepting the article for publication acknowledges, that the United States Government retains a non-exclusive, paid-up, irrevocable, world-wide license to publish or reproduce the published form of this manuscript, or allow others to do so, for United States Government purposes.

\bibliography{bibliography}

\begin{thebibliography}{23}%
\makeatletter
\providecommand \@ifxundefined [1]{%
 \@ifx{#1\undefined}
}%
\providecommand \@ifnum [1]{%
 \ifnum #1\expandafter \@firstoftwo
 \else \expandafter \@secondoftwo
 \fi
}%
\providecommand \@ifx [1]{%
 \ifx #1\expandafter \@firstoftwo
 \else \expandafter \@secondoftwo
 \fi
}%
\providecommand \natexlab [1]{#1}%
\providecommand \enquote  [1]{``#1''}%
\providecommand \bibnamefont  [1]{#1}%
\providecommand \bibfnamefont [1]{#1}%
\providecommand \citenamefont [1]{#1}%
\providecommand \href@noop [0]{\@secondoftwo}%
\providecommand \href [0]{\begingroup \@sanitize@url \@href}%
\providecommand \@href[1]{\@@startlink{#1}\@@href}%
\providecommand \@@href[1]{\endgroup#1\@@endlink}%
\providecommand \@sanitize@url [0]{\catcode `\\12\catcode `\$12\catcode
  `\&12\catcode `\#12\catcode `\^12\catcode `\_12\catcode `\%12\relax}%
\providecommand \@@startlink[1]{}%
\providecommand \@@endlink[0]{}%
\providecommand \url  [0]{\begingroup\@sanitize@url \@url }%
\providecommand \@url [1]{\endgroup\@href {#1}{\urlprefix }}%
\providecommand \urlprefix  [0]{URL }%
\providecommand \Eprint [0]{\href }%
\providecommand \doibase [0]{http://dx.doi.org/}%
\providecommand \selectlanguage [0]{\@gobble}%
\providecommand \bibinfo  [0]{\@secondoftwo}%
\providecommand \bibfield  [0]{\@secondoftwo}%
\providecommand \translation [1]{[#1]}%
\providecommand \BibitemOpen [0]{}%
\providecommand \bibitemStop [0]{}%
\providecommand \bibitemNoStop [0]{.\EOS\space}%
\providecommand \EOS [0]{\spacefactor3000\relax}%
\providecommand \BibitemShut  [1]{\csname bibitem#1\endcsname}%
\let\auto@bib@innerbib\@empty
\bibitem [{\citenamefont {Brizard}\ and\ \citenamefont
  {Hahm}(2007)}]{Brizard07}%
  \BibitemOpen
  \bibfield  {author} {\bibinfo {author} {\bibfnamefont {A.~J.}\ \bibnamefont
  {Brizard}}\ and\ \bibinfo {author} {\bibfnamefont {T.~S.}\ \bibnamefont
  {Hahm}},\ }\href {\doibase 10.1103/RevModPhys.79.421} {\bibfield  {journal}
  {\bibinfo  {journal} {Rev. Mod. Phys.}\ }\textbf {\bibinfo {volume} {79}},\
  \bibinfo {pages} {421} (\bibinfo {year} {2007})}\BibitemShut {NoStop}%
\bibitem [{\citenamefont {Lee}(1987)}]{Lee87}%
  \BibitemOpen
  \bibfield  {author} {\bibinfo {author} {\bibfnamefont {W.}~\bibnamefont
  {Lee}},\ }\href@noop {} {\bibfield  {journal} {\bibinfo  {journal} {Journal
  of Computational Physics}\ }\textbf {\bibinfo {volume} {72}},\ \bibinfo
  {pages} {243 } (\bibinfo {year} {1987})}\BibitemShut {NoStop}%
\bibitem [{\citenamefont {Mishchenko}\ \emph {et~al.}(2005)\citenamefont
  {Mishchenko}, \citenamefont {K\"onies},\ and\ \citenamefont
  {Hatzky}}]{Mishchenko05}%
  \BibitemOpen
  \bibfield  {author} {\bibinfo {author} {\bibfnamefont {A.}~\bibnamefont
  {Mishchenko}}, \bibinfo {author} {\bibfnamefont {A.}~\bibnamefont
  {K\"onies}}, \ and\ \bibinfo {author} {\bibfnamefont {R.}~\bibnamefont
  {Hatzky}},\ }\href@noop {} {\bibfield  {journal} {\bibinfo  {journal}
  {Physics of Plasmas}\ }\textbf {\bibinfo {volume} {12}},\ \bibinfo {eid}
  {062305} (\bibinfo {year} {2005})}\BibitemShut {NoStop}%
\bibitem [{\citenamefont {Ku}\ \emph {et~al.}(2016)\citenamefont {Ku},
  \citenamefont {Hager}, \citenamefont {Chang}, \citenamefont {Kwon},\ and\
  \citenamefont {Parker}}]{Ku16}%
  \BibitemOpen
  \bibfield  {author} {\bibinfo {author} {\bibfnamefont {S.}~\bibnamefont
  {Ku}}, \bibinfo {author} {\bibfnamefont {R.}~\bibnamefont {Hager}}, \bibinfo
  {author} {\bibfnamefont {C.}~\bibnamefont {Chang}}, \bibinfo {author}
  {\bibfnamefont {J.}~\bibnamefont {Kwon}}, \ and\ \bibinfo {author}
  {\bibfnamefont {S.}~\bibnamefont {Parker}},\ }\href@noop {} {\bibfield
  {journal} {\bibinfo  {journal} {Journal of Computational Physics}\ }\textbf
  {\bibinfo {volume} {315}},\ \bibinfo {pages} {467 } (\bibinfo {year}
  {2016})}\BibitemShut {NoStop}%
\bibitem [{\citenamefont {Ku}\ \emph {et~al.}()\citenamefont {Ku},
  \citenamefont {Chang}, \citenamefont {Hager}, \citenamefont {Churchill},
  \citenamefont {Tyna}, \citenamefont {Cziegler†}, \citenamefont {Greenwald},
  \citenamefont {Hughes}, \citenamefont {Parker}, \citenamefont {Adams},
  \citenamefont {D’Azevedo},\ and\ \citenamefont {Worley}}]{Ku18}%
  \BibitemOpen
  \bibfield  {author} {\bibinfo {author} {\bibfnamefont {S.}~\bibnamefont
  {Ku}}, \bibinfo {author} {\bibfnamefont {C.}~\bibnamefont {Chang}}, \bibinfo
  {author} {\bibfnamefont {R.}~\bibnamefont {Hager}}, \bibinfo {author}
  {\bibfnamefont {R.}~\bibnamefont {Churchill}}, \bibinfo {author}
  {\bibfnamefont {G.}~\bibnamefont {Tyna}}, \bibinfo {author} {\bibfnamefont
  {I.}~\bibnamefont {Cziegler†}}, \bibinfo {author} {\bibfnamefont
  {M.}~\bibnamefont {Greenwald}}, \bibinfo {author} {\bibfnamefont
  {J.}~\bibnamefont {Hughes}}, \bibinfo {author} {\bibfnamefont {S.~E.}\
  \bibnamefont {Parker}}, \bibinfo {author} {\bibfnamefont {M.}~\bibnamefont
  {Adams}}, \bibinfo {author} {\bibfnamefont {E.}~\bibnamefont {D’Azevedo}},
  \ and\ \bibinfo {author} {\bibfnamefont {P.}~\bibnamefont {Worley}},\
  }\href@noop {} {\bibinfo  {journal} {submitted to Physics of Plasmas}\
  }\BibitemShut {NoStop}%
\bibitem [{\citenamefont {Tran}\ \emph {et~al.}(1999)\citenamefont {Tran},
  \citenamefont {Appert}, \citenamefont {Fivaz}, \citenamefont {Jost},
  \citenamefont {Vaclavik},\ and\ \citenamefont {Villard}}]{Tran99}%
  \BibitemOpen
\bibfield  {journal} {  }\bibfield  {author} {\bibinfo {author} {\bibfnamefont
  {T.}~\bibnamefont {Tran}}, \bibinfo {author} {\bibfnamefont {K.}~\bibnamefont
  {Appert}}, \bibinfo {author} {\bibfnamefont {M.}~\bibnamefont {Fivaz}},
  \bibinfo {author} {\bibfnamefont {G.}~\bibnamefont {Jost}}, \bibinfo {author}
  {\bibfnamefont {J.}~\bibnamefont {Vaclavik}}, \ and\ \bibinfo {author}
  {\bibfnamefont {L.}~\bibnamefont {Villard}},\ }\href@noop {} {\bibfield
  {journal} {\bibinfo  {journal} {Theory of Fusion Plasmas, Int. Workshop
  (Bologna Editrice Compositori,Societ\`a Italiana di Fisica)}\ ,\ \bibinfo
  {pages} {45}} (\bibinfo {year} {1999})}\BibitemShut {NoStop}%
\bibitem [{\citenamefont {Chen}\ and\ \citenamefont {Parker}(2003)}]{Parker03}%
  \BibitemOpen
  \bibfield  {author} {\bibinfo {author} {\bibfnamefont {Y.}~\bibnamefont
  {Chen}}\ and\ \bibinfo {author} {\bibfnamefont {S.~E.}\ \bibnamefont
  {Parker}},\ }\href {\doibase https://doi.org/10.1016/S0021-9991(03)00228-6}
  {\bibfield  {journal} {\bibinfo  {journal} {Journal of Computational
  Physics}\ }\textbf {\bibinfo {volume} {189}},\ \bibinfo {pages} {463 }
  (\bibinfo {year} {2003})}\BibitemShut {NoStop}%
\bibitem [{\citenamefont {Idomura}\ \emph {et~al.}(2008)\citenamefont
  {Idomura}, \citenamefont {Ida}, \citenamefont {Kano}, \citenamefont {Aiba},\
  and\ \citenamefont {Tokuda}}]{Idomura08}%
  \BibitemOpen
  \bibfield  {author} {\bibinfo {author} {\bibfnamefont {Y.}~\bibnamefont
  {Idomura}}, \bibinfo {author} {\bibfnamefont {M.}~\bibnamefont {Ida}},
  \bibinfo {author} {\bibfnamefont {T.}~\bibnamefont {Kano}}, \bibinfo {author}
  {\bibfnamefont {N.}~\bibnamefont {Aiba}}, \ and\ \bibinfo {author}
  {\bibfnamefont {S.}~\bibnamefont {Tokuda}},\ }\href {\doibase
  http://dx.doi.org/10.1016/j.cpc.2008.04.005} {\bibfield  {journal} {\bibinfo
  {journal} {Computer Physics Communications}\ }\textbf {\bibinfo {volume}
  {179}},\ \bibinfo {pages} {391 } (\bibinfo {year} {2008})}\BibitemShut
  {NoStop}%
\bibitem [{\citenamefont {Lin}\ and\ \citenamefont {Lee}(1995)}]{Lin95}%
  \BibitemOpen
  \bibfield  {author} {\bibinfo {author} {\bibfnamefont {Z.}~\bibnamefont
  {Lin}}\ and\ \bibinfo {author} {\bibfnamefont {W.~W.}\ \bibnamefont {Lee}},\
  }\href@noop {} {\bibfield  {journal} {\bibinfo  {journal} {Phys. Rev. E}\
  }\textbf {\bibinfo {volume} {52}},\ \bibinfo {pages} {5646} (\bibinfo {year}
  {1995})}\BibitemShut {NoStop}%
\bibitem [{\citenamefont {Jenko}\ \emph {et~al.}(2000)\citenamefont {Jenko},
  \citenamefont {Dorland}, \citenamefont {Kotschenreuther},\ and\ \citenamefont
  {Rogers}}]{Jenko00}%
  \BibitemOpen
  \bibfield  {author} {\bibinfo {author} {\bibfnamefont {F.}~\bibnamefont
  {Jenko}}, \bibinfo {author} {\bibfnamefont {W.}~\bibnamefont {Dorland}},
  \bibinfo {author} {\bibfnamefont {M.}~\bibnamefont {Kotschenreuther}}, \ and\
  \bibinfo {author} {\bibfnamefont {B.~N.}\ \bibnamefont {Rogers}},\ }\href
  {\doibase 10.1063/1.874014} {\bibfield  {journal} {\bibinfo  {journal}
  {Physics of Plasmas}\ }\textbf {\bibinfo {volume} {7}},\ \bibinfo {pages}
  {1904} (\bibinfo {year} {2000})}\BibitemShut {NoStop}%
\bibitem [{\citenamefont {G\"orler}\ \emph {et~al.}(2011)\citenamefont
  {G\"orler}, \citenamefont {Lapillonne}, \citenamefont {Brunner},
  \citenamefont {Dannert}, \citenamefont {Jenko}, \citenamefont {Merz},\ and\
  \citenamefont {Told}}]{Goerler11}%
  \BibitemOpen
  \bibfield  {author} {\bibinfo {author} {\bibfnamefont {T.}~\bibnamefont
  {G\"orler}}, \bibinfo {author} {\bibfnamefont {X.}~\bibnamefont
  {Lapillonne}}, \bibinfo {author} {\bibfnamefont {S.}~\bibnamefont {Brunner}},
  \bibinfo {author} {\bibfnamefont {T.}~\bibnamefont {Dannert}}, \bibinfo
  {author} {\bibfnamefont {F.}~\bibnamefont {Jenko}}, \bibinfo {author}
  {\bibfnamefont {F.}~\bibnamefont {Merz}}, \ and\ \bibinfo {author}
  {\bibfnamefont {D.}~\bibnamefont {Told}},\ }\href {\doibase
  10.1016/j.jcp.2011.05.034} {\bibfield  {journal} {\bibinfo  {journal}
  {Journal of Computational Physics}\ }\textbf {\bibinfo {volume} {230}},\
  \bibinfo {pages} {7053 } (\bibinfo {year} {2011})}\BibitemShut {NoStop}%
\bibitem [{\citenamefont {Jarema}\ \emph {et~al.}(2017)\citenamefont {Jarema},
  \citenamefont {Bungartz}, \citenamefont {Görler}, \citenamefont {Jenko},
  \citenamefont {Neckel},\ and\ \citenamefont {Told}}]{Jarema17}%
  \BibitemOpen
  \bibfield  {author} {\bibinfo {author} {\bibfnamefont {D.}~\bibnamefont
  {Jarema}}, \bibinfo {author} {\bibfnamefont {H.}~\bibnamefont {Bungartz}},
  \bibinfo {author} {\bibfnamefont {T.}~\bibnamefont {Görler}}, \bibinfo
  {author} {\bibfnamefont {F.}~\bibnamefont {Jenko}}, \bibinfo {author}
  {\bibfnamefont {T.}~\bibnamefont {Neckel}}, \ and\ \bibinfo {author}
  {\bibfnamefont {D.}~\bibnamefont {Told}},\ }\href@noop {} {\bibfield
  {journal} {\bibinfo  {journal} {Computer Physics Communications}\ }\textbf
  {\bibinfo {volume} {215}},\ \bibinfo {pages} {49 } (\bibinfo {year}
  {2017})}\BibitemShut {NoStop}%
\bibitem [{\citenamefont {Grandgirard}\ \emph {et~al.}(2007)\citenamefont
  {Grandgirard}, \citenamefont {Sarazin}, \citenamefont {Angelino},
  \citenamefont {Bottino}, \citenamefont {Crouseilles}, \citenamefont {Darmet},
  \citenamefont {Dif-Pradalier}, \citenamefont {Garbet}, \citenamefont
  {Ghendrih}, \citenamefont {Jolliet}, \citenamefont {Latu}, \citenamefont
  {Sonnendrücker},\ and\ \citenamefont {Villard}}]{Grandgirard07}%
  \BibitemOpen
  \bibfield  {author} {\bibinfo {author} {\bibfnamefont {V.}~\bibnamefont
  {Grandgirard}}, \bibinfo {author} {\bibfnamefont {Y.}~\bibnamefont
  {Sarazin}}, \bibinfo {author} {\bibfnamefont {P.}~\bibnamefont {Angelino}},
  \bibinfo {author} {\bibfnamefont {A.}~\bibnamefont {Bottino}}, \bibinfo
  {author} {\bibfnamefont {N.}~\bibnamefont {Crouseilles}}, \bibinfo {author}
  {\bibfnamefont {G.}~\bibnamefont {Darmet}}, \bibinfo {author} {\bibfnamefont
  {G.}~\bibnamefont {Dif-Pradalier}}, \bibinfo {author} {\bibfnamefont
  {X.}~\bibnamefont {Garbet}}, \bibinfo {author} {\bibfnamefont
  {P.}~\bibnamefont {Ghendrih}}, \bibinfo {author} {\bibfnamefont
  {S.}~\bibnamefont {Jolliet}}, \bibinfo {author} {\bibfnamefont
  {G.}~\bibnamefont {Latu}}, \bibinfo {author} {\bibfnamefont {E.}~\bibnamefont
  {Sonnendrücker}}, \ and\ \bibinfo {author} {\bibfnamefont {L.}~\bibnamefont
  {Villard}},\ }\href@noop {} {\bibfield  {journal} {\bibinfo  {journal}
  {Plasma Physics and Controlled Fusion}\ }\textbf {\bibinfo {volume} {49}},\
  \bibinfo {pages} {B173} (\bibinfo {year} {2007})}\BibitemShut {NoStop}%
\bibitem [{\citenamefont {Steiner}\ \emph {et~al.}(2015)\citenamefont
  {Steiner}, \citenamefont {Mehrenberger}, \citenamefont {Crouseilles},
  \citenamefont {Grandgirard}, \citenamefont {Latu},\ and\ \citenamefont
  {Rozar}}]{Steiner15}%
  \BibitemOpen
  \bibfield  {author} {\bibinfo {author} {\bibfnamefont {C.}~\bibnamefont
  {Steiner}}, \bibinfo {author} {\bibfnamefont {M.}~\bibnamefont
  {Mehrenberger}}, \bibinfo {author} {\bibfnamefont {N.}~\bibnamefont
  {Crouseilles}}, \bibinfo {author} {\bibfnamefont {V.}~\bibnamefont
  {Grandgirard}}, \bibinfo {author} {\bibfnamefont {G.}~\bibnamefont {Latu}}, \
  and\ \bibinfo {author} {\bibfnamefont {F.}~\bibnamefont {Rozar}},\
  }\href@noop {} {\bibfield  {journal} {\bibinfo  {journal} {The European
  Physical Journal D}\ }\textbf {\bibinfo {volume} {69}},\ \bibinfo {pages}
  {18} (\bibinfo {year} {2015})}\BibitemShut {NoStop}%
\bibitem [{\citenamefont {{Rozar, Fabien}}\ \emph {et~al.}(2016)\citenamefont
  {{Rozar, Fabien}}, \citenamefont {{Steiner, Christophe}}, \citenamefont
  {{Latu, Guillaume}}, \citenamefont {{Mehrenberger, Michel}}, \citenamefont
  {{Grandgirard, Virginie}}, \citenamefont {{Bigot, Julien}}, \citenamefont
  {{Cartier-Michaud, Thomas}},\ and\ \citenamefont {{Roman, Jean}}}]{Rozar16}%
  \BibitemOpen
  \bibfield  {author} {\bibinfo {author} {\bibnamefont {{Rozar, Fabien}}},
  \bibinfo {author} {\bibnamefont {{Steiner, Christophe}}}, \bibinfo {author}
  {\bibnamefont {{Latu, Guillaume}}}, \bibinfo {author} {\bibnamefont
  {{Mehrenberger, Michel}}}, \bibinfo {author} {\bibnamefont {{Grandgirard,
  Virginie}}}, \bibinfo {author} {\bibnamefont {{Bigot, Julien}}}, \bibinfo
  {author} {\bibnamefont {{Cartier-Michaud, Thomas}}}, \ and\ \bibinfo {author}
  {\bibnamefont {{Roman, Jean}}},\ }\href@noop {} {\bibfield  {journal}
  {\bibinfo  {journal} {ESAIM: Proc.}\ }\textbf {\bibinfo {volume} {53}},\
  \bibinfo {pages} {191} (\bibinfo {year} {2016})}\BibitemShut {NoStop}%
\bibitem [{\citenamefont {Candy}\ and\ \citenamefont {Waltz}(2003)}]{Candy03}%
  \BibitemOpen
  \bibfield  {author} {\bibinfo {author} {\bibfnamefont {J.}~\bibnamefont
  {Candy}}\ and\ \bibinfo {author} {\bibfnamefont {R.}~\bibnamefont {Waltz}},\
  }\href {\doibase https://doi.org/10.1016/S0021-9991(03)00079-2} {\bibfield
  {journal} {\bibinfo  {journal} {Journal of Computational Physics}\ }\textbf
  {\bibinfo {volume} {186}},\ \bibinfo {pages} {545 } (\bibinfo {year}
  {2003})}\BibitemShut {NoStop}%
\bibitem [{\citenamefont {Dominski}\ \emph {et~al.}(2015)\citenamefont
  {Dominski}, \citenamefont {Brunner}, \citenamefont {G\"orler}, \citenamefont
  {Jenko}, \citenamefont {Told},\ and\ \citenamefont {Villard}}]{dominski15}%
  \BibitemOpen
  \bibfield  {author} {\bibinfo {author} {\bibfnamefont {J.}~\bibnamefont
  {Dominski}}, \bibinfo {author} {\bibfnamefont {S.}~\bibnamefont {Brunner}},
  \bibinfo {author} {\bibfnamefont {T.}~\bibnamefont {G\"orler}}, \bibinfo
  {author} {\bibfnamefont {F.}~\bibnamefont {Jenko}}, \bibinfo {author}
  {\bibfnamefont {D.}~\bibnamefont {Told}}, \ and\ \bibinfo {author}
  {\bibfnamefont {L.}~\bibnamefont {Villard}},\ }\href@noop {} {\bibfield
  {journal} {\bibinfo  {journal} {Physics of Plasmas}\ }\textbf {\bibinfo
  {volume} {22}},\ \bibinfo {eid} {062303} (\bibinfo {year}
  {2015})}\BibitemShut {NoStop}%
\bibitem [{\citenamefont {Dominski}\ \emph {et~al.}(2017)\citenamefont
  {Dominski}, \citenamefont {McMillan}, \citenamefont {Brunner}, \citenamefont
  {Merlo}, \citenamefont {Tran},\ and\ \citenamefont {Villard}}]{dominski17}%
  \BibitemOpen
  \bibfield  {author} {\bibinfo {author} {\bibfnamefont {J.}~\bibnamefont
  {Dominski}}, \bibinfo {author} {\bibfnamefont {B.~F.}\ \bibnamefont
  {McMillan}}, \bibinfo {author} {\bibfnamefont {S.}~\bibnamefont {Brunner}},
  \bibinfo {author} {\bibfnamefont {G.}~\bibnamefont {Merlo}}, \bibinfo
  {author} {\bibfnamefont {T.-M.}\ \bibnamefont {Tran}}, \ and\ \bibinfo
  {author} {\bibfnamefont {L.}~\bibnamefont {Villard}},\ }\href@noop {}
  {\bibfield  {journal} {\bibinfo  {journal} {Physics of Plasmas}\ }\textbf
  {\bibinfo {volume} {24}},\ \bibinfo {pages} {022308} (\bibinfo {year}
  {2017})}\BibitemShut {NoStop}%
\bibitem [{\citenamefont {Howard}\ \emph {et~al.}(2014)\citenamefont {Howard},
  \citenamefont {Holland}, \citenamefont {White}, \citenamefont {Greenwald},\
  and\ \citenamefont {Candy}}]{Howard14}%
  \BibitemOpen
  \bibfield  {author} {\bibinfo {author} {\bibfnamefont {N.~T.}\ \bibnamefont
  {Howard}}, \bibinfo {author} {\bibfnamefont {C.}~\bibnamefont {Holland}},
  \bibinfo {author} {\bibfnamefont {A.~E.}\ \bibnamefont {White}}, \bibinfo
  {author} {\bibfnamefont {M.}~\bibnamefont {Greenwald}}, \ and\ \bibinfo
  {author} {\bibfnamefont {J.}~\bibnamefont {Candy}},\ }\href@noop {}
  {\bibfield  {journal} {\bibinfo  {journal} {Physics of Plasmas}\ }\textbf
  {\bibinfo {volume} {21}},\ \bibinfo {eid} {112510} (\bibinfo {year}
  {2014})}\BibitemShut {NoStop}%
\bibitem [{\citenamefont {Maeyama}\ \emph {et~al.}(2015)\citenamefont
  {Maeyama}, \citenamefont {Idomura}, \citenamefont {Watanabe}, \citenamefont
  {Nakata}, \citenamefont {Yagi}, \citenamefont {Miyato}, \citenamefont
  {Ishizawa},\ and\ \citenamefont {Nunami}}]{Maeyama15}%
  \BibitemOpen
  \bibfield  {author} {\bibinfo {author} {\bibfnamefont {S.}~\bibnamefont
  {Maeyama}}, \bibinfo {author} {\bibfnamefont {Y.}~\bibnamefont {Idomura}},
  \bibinfo {author} {\bibfnamefont {T.-H.}\ \bibnamefont {Watanabe}}, \bibinfo
  {author} {\bibfnamefont {M.}~\bibnamefont {Nakata}}, \bibinfo {author}
  {\bibfnamefont {M.}~\bibnamefont {Yagi}}, \bibinfo {author} {\bibfnamefont
  {N.}~\bibnamefont {Miyato}}, \bibinfo {author} {\bibfnamefont
  {A.}~\bibnamefont {Ishizawa}}, \ and\ \bibinfo {author} {\bibfnamefont
  {M.}~\bibnamefont {Nunami}},\ }\href@noop {} {\bibfield  {journal} {\bibinfo
  {journal} {Phys. Rev. Lett.}\ }\textbf {\bibinfo {volume} {114}},\ \bibinfo
  {pages} {255002} (\bibinfo {year} {2015})}\BibitemShut {NoStop}%
\bibitem [{\citenamefont {Fivaz}\ \emph {et~al.}(1998)\citenamefont {Fivaz},
  \citenamefont {Brunner}, \citenamefont {de~Ridder}, \citenamefont {Sauter},
  \citenamefont {Tran}, \citenamefont {Vaclavik}, \citenamefont {Villard},\
  and\ \citenamefont {Appert}}]{Fivaz98}%
  \BibitemOpen
  \bibfield  {author} {\bibinfo {author} {\bibfnamefont {M.}~\bibnamefont
  {Fivaz}}, \bibinfo {author} {\bibfnamefont {S.}~\bibnamefont {Brunner}},
  \bibinfo {author} {\bibfnamefont {G.}~\bibnamefont {de~Ridder}}, \bibinfo
  {author} {\bibfnamefont {O.}~\bibnamefont {Sauter}}, \bibinfo {author}
  {\bibfnamefont {T.}~\bibnamefont {Tran}}, \bibinfo {author} {\bibfnamefont
  {J.}~\bibnamefont {Vaclavik}}, \bibinfo {author} {\bibfnamefont
  {L.}~\bibnamefont {Villard}}, \ and\ \bibinfo {author} {\bibfnamefont
  {K.}~\bibnamefont {Appert}},\ }\href@noop {} {\bibfield  {journal} {\bibinfo
  {journal} {Computer Physics Communications}\ }\textbf {\bibinfo {volume}
  {111}},\ \bibinfo {pages} {27 } (\bibinfo {year} {1998})}\BibitemShut
  {NoStop}%
\bibitem [{\citenamefont {Lapillonne}\ \emph {et~al.}(2009)\citenamefont
  {Lapillonne}, \citenamefont {Brunner}, \citenamefont {Dannert}, \citenamefont
  {Jolliet}, \citenamefont {Marinoni}, \citenamefont {Villard}, \citenamefont
  {G\"orler}, \citenamefont {Jenko},\ and\ \citenamefont {Merz}}]{Lapillonne}%
  \BibitemOpen
  \bibfield  {author} {\bibinfo {author} {\bibfnamefont {X.}~\bibnamefont
  {Lapillonne}}, \bibinfo {author} {\bibfnamefont {S.}~\bibnamefont {Brunner}},
  \bibinfo {author} {\bibfnamefont {T.}~\bibnamefont {Dannert}}, \bibinfo
  {author} {\bibfnamefont {S.}~\bibnamefont {Jolliet}}, \bibinfo {author}
  {\bibfnamefont {A.}~\bibnamefont {Marinoni}}, \bibinfo {author}
  {\bibfnamefont {L.}~\bibnamefont {Villard}}, \bibinfo {author} {\bibfnamefont
  {T.}~\bibnamefont {G\"orler}}, \bibinfo {author} {\bibfnamefont
  {F.}~\bibnamefont {Jenko}}, \ and\ \bibinfo {author} {\bibfnamefont
  {F.}~\bibnamefont {Merz}},\ }\href {\doibase 10.1063/1.3096710} {\bibfield
  {journal} {\bibinfo  {journal} {Physics of Plasmas}\ }\textbf {\bibinfo
  {volume} {16}},\ \bibinfo {eid} {032308} (\bibinfo {year}
  {2009})}\BibitemShut {NoStop}%
\bibitem [{\citenamefont {Burckel}\ \emph {et~al.}(2010)\citenamefont
  {Burckel}, \citenamefont {Sauter}, \citenamefont {Angioni}, \citenamefont
  {Candy}, \citenamefont {Fable},\ and\ \citenamefont
  {Lapillonne}}]{Burckel10}%
  \BibitemOpen
  \bibfield  {author} {\bibinfo {author} {\bibfnamefont {A.}~\bibnamefont
  {Burckel}}, \bibinfo {author} {\bibfnamefont {O.}~\bibnamefont {Sauter}},
  \bibinfo {author} {\bibfnamefont {C.}~\bibnamefont {Angioni}}, \bibinfo
  {author} {\bibfnamefont {J.}~\bibnamefont {Candy}}, \bibinfo {author}
  {\bibfnamefont {E.}~\bibnamefont {Fable}}, \ and\ \bibinfo {author}
  {\bibfnamefont {X.}~\bibnamefont {Lapillonne}},\ }\href@noop {} {\bibfield
  {journal} {\bibinfo  {journal} {Journal of Physics: Conference Series}\
  }\textbf {\bibinfo {volume} {260}},\ \bibinfo {pages} {012006} (\bibinfo
  {year} {2010})}\BibitemShut {NoStop}%
\end{thebibliography}%

\end{document}